\documentclass[moor]{informs1_hide}              

\usepackage{natbib}
 \NatBibNumeric
 \def\bibfont{\small}%
 \bibpunct[, ]{[}{]}{,}{n}{}{,}%

\usepackage[colorlinks=true,breaklinks=true,bookmarks=true,urlcolor=blue,
     citecolor=blue,linkcolor=blue,bookmarksopen=false,draft=false]{hyperref}

\def\EMAIL#1{\href{mailto:#1}{#1}}
\def\URL#1{\href{#1}{#1}}         

\usepackage[dvipsnames]{xcolor}
\usepackage{enumitem}
\setlist[enumerate]{leftmargin=.5in}
\setlist[itemize]{leftmargin=.5in}

\usepackage{amsmath,amssymb} 
\usepackage[capitalize]{cleveref}
\crefname{subsection}{subsection}{subsections}
\crefname{algocf}{algorithm}{algorithms}
\usepackage[normalem]{ulem}
\usepackage{bm}

\usepackage{algorithm}
\usepackage{algpseudocode}

\newcommand{\set}[1]{\{ #1 \}}
\newcommand{\bE}{\mathbb{E}}

\newcommand{\bR}{\mathbb{R}}
\newcommand{\ALG}{\mathsf{ALG}}
\newcommand{\LP}{\mathsf{LP}}
\newcommand{\OPT}{\mathsf{OPT}}
\newcommand{\euler}{\mathsf{e}}
\newcommand{\eps}{\varepsilon}
\newcommand{\patience}{t}
\newcommand{\pat}{\patience}
\newcommand{\Pois}{\mathrm{Pois}}
\newcommand{\til}[1]{\widetilde{#1}}

\TheoremsNumberedThrough     

\EquationsNumberedThrough    

\newtheorem{result}{Result}

\NewDocumentEnvironment{myproof}{o}
{\IfNoValueTF{#1}{\paragraph{{Proof.} }} {\paragraph{{#1.} }} }
{\hfill$\Halmos$\par}

\newcommand{\revisionchange}[1]{{\leavevmode #1}}

\begin{document}




\TITLE{Improved Guarantees for Offline Stochastic Matching via New Ordered Contention Resolution Schemes}

\ARTICLEAUTHORS{%
\AUTHOR{Brian Brubach}
\AFF{Computer Science Department, Wellesley College, \EMAIL{bb100@wellesley.edu}, \URL{}}
\AUTHOR{Nathaniel Grammel}
\AFF{Department of Computer Science, University of Maryland, College Park, \EMAIL{ngrammel@umd.edu}, \URL{}}
\AUTHOR{Will Ma}
\AFF{Graduate School of Business, Columbia University, New York, \EMAIL{wm2428@gsb.columbia.edu}, \URL{}}
\AUTHOR{Calum MacRury}
\AFF{School of Industrial and Systems Engineering (ISyE), Georgia Tech, Atlanta,
\EMAIL{calum.macrury@isye.gatech.edu}} \URL{}

\AUTHOR{Aravind Srinivasan}
\AFF{Department of Computer Science, University of Maryland, College Park, \EMAIL{asriniv1@umd.edu}, \URL{}}
} 

\ABSTRACT{%
  Matching is one of the most fundamental and broadly applicable problems across many domains. In these diverse real-world applications, there is often a degree of uncertainty in the input which has led to the study of stochastic matching models. Here, each edge in the graph has a known, independent probability of existing derived from some prediction. Algorithms must probe edges to determine existence and match them irrevocably if they exist. Further, each vertex may have a patience constraint denoting how many of its neighboring edges can be probed. We present new ordered contention resolution schemes yielding improved approximation guarantees for some of the foundational problems studied in this area. For stochastic matching with patience constraints in general graphs, we provide a $0.382$-approximate algorithm assuming each vertex has patience at least $2$.
  Under this assumption, we improve upon the previous best $0.31$-approximation of Baveja et al.~(2018). When the vertices do not have patience constraints, we describe a $0.432$-approximate random order probing algorithm with several corollaries such as an improved guarantee for the Prophet Secretary problem under Edge Arrivals. Finally, for the special case of bipartite graphs with unit patience constraints on one of the partitions, we show a $0.632$-approximate algorithm that improves on the recent $1/3$-guarantee of Hikima et al.~(2021).
}%



\maketitle

%


\section{Introduction}
\label{sec:intro}


The offline stochastic matching problem is about finding a maximum matching on a weighted graph.
However, each edge $e$ is \textit{active} independently according to a known probability $p_e$, and only active edges can be matched.
The set of active edges is initially unknown.
An edge whose endpoints are unmatched can be \textit{probed} to determine whether it is active, and if so, it is irrevocably inserted into the matching.
The objective is to sequentially probe the edges in a way to maximize the expected weighted matching at the end.

Matching problems arise in numerous \revisionchange{deployed systems}, especially those dealing with allocation and scheduling. See, for example, the works of \citet{DBLP:conf/ijcai/AhmadiADFK20,DBLP:conf/ijcai/AhmedDF17,brubach2020attenuate,brubach21FYS,baveja2018improved} for applications to advertising, e-commerce, organ exchange, online dating, peer review, school matching, and hiring; and \citet{hikima2021integrated,DBLP:conf/aaai/Nanda0SDS20,DBLP:conf/aaai/XuSCDSSTT19} for applications to ride sharing, crowdsourcing (worker-task assignment), and recommendation systems. The work of \citet{esfandiari2016neuripsmatching,antoniadis2020neuripsmatching} gives further discussion on applications to e-commerce and internet advertising, and illustrates the importance of designing algorithms with good theoretical \revisionchange{guarantees.} 

Stochastic edges play a major role in properly capturing many of these applications when we account for hidden information about whether a proposed match will be accepted. Is a kidney donor a good match? Will a worker accept an offered task? Does a user want to click on this ad? The uncertainty inherent in the presence or absence of various edges naturally leads to the need for stochastic optimization. Achieving good results when the edge probabilities are known demonstrates the value of learning distributions in AI. 

We contribute improved algorithms for a range of fundamental problems, as well as useful-special-case problems, in the realm of stochastic matching. Although our results are phrased for the offline setting, some of our algorithms also translate to the \emph{random-order} online setting in which the edges arrive in random order. This online arrival model is well-motivated in \revisionchange{these settings} due to the inability to control the order in which the agents arrive.

Below, we further describe three typical features that may be present in stochastic matching problems. 

\textbf{Patience constraints.}
The offline stochastic matching problem is typically considered with the input including an integral \emph{patience parameter} $t_v$ for each vertex $v$ known as a \emph{patience constraint} or timeout (see, e.g., \citet{chen2009approximatingMatches}, who first introduced this formulation of the problem).
This adds the constraint that at most $t_v$ of the edges incident to $v$ can be probed,
which is motivated by applications in which the vertices represent users who are only willing to view a finite number of potential matches.
A vertex $v$ which has been probed $t_v$ times is said to have \textit{timed out}.
To capture the absence of patience constraints in some models, we allow patience parameters to be $\infty$.

\textbf{Probing orders allowed.}
Most generally, we allow algorithms to probe the edges in any order based on the realizations of past probes.
Sometimes we impose that the algorithm must make a single pass through the edges, according to a uniform random permutation.
This can also represent an ``online'' setting in which the edges arrive in a uniformly random order, and upon arrival, an edge must either be probed (only possible if both endpoints are unmatched and have remaining patience) or irrevocably discarded. 
We say that such algorithms are \textit{random-order}.

\textbf{Special case of graphs considered.}
We introduce a new class of graphs, \textit{bipartite graphs with a unit-patience} side, for which we derive further improved guarantees.
Such a graph can be divided into two sides such that: (1) all edges are between vertices on different sides; and (2) for one of the sides, all patiences are 1.
This subclass of graphs captures a problem of interest in AI with applications such as crowdsourcing and ride-hailing, described in \citet{hikima2021integrated}.

\subsection{Summary of Results} \label{sec:introResults}

To derive computationally-efficient probing algorithms, we address the standard approach of first solving a fractional relaxation, which prescribes a probability $y_e\in[0,1]$ with which each edge $e$ should be probed.
These relaxed values $y_e$ only have to satisfy the matching and patience constraints in expectation, while a probing algorithm must satisfy these constraints with probability (w.p.) $1$.
However, we can still hope to probe every edge $e$ with probability at least $c\cdot y_e$, for some $c\le1$.
It is well-known that such an algorithm would then be $c$-\textit{approximate}, i.e.\ its expected weight matched would be at least $c$ times that of an optimal probing algorithm.


\begin{result}\label{result:382}
A \textbf{0.382}-approximate \textit{random-order} probing algorithm for general graphs with patience at least $2$.
\end{result}
We note that the previous version of the paper incorrectly claimed this result held for all patience values. However, on graphs with patience values of $1$, there exists an input where our algorithm fails to exceed $1/3$ against the fractional relaxation we consider (see \Cref{rem:unit_patience} from \Cref{sec:analysis_of_algorithm} for details). Restricted to inputs
with patience at least $2$, \Cref{result:382} improves on the \textbf{0.25}-approximate algorithm of \citet{bansal2012lp}, the \textbf{0.269}-approximate algorithm of \citet{adamczyk2015improved}, and the \textbf{0.31}-approximate algorithm of \citet{baveja2018improved}. Since our paper first appeared, the state-of-the-art for general graphs is now due to \citet{DBLP:journals/corr/abs-2205-08667}, who attained a $\textbf{0.395}$-approximate algorithm. For the special case of bipartite graphs, \citet{derakhshan2026approximation} recently attained a
$\textbf{0.58}$-approximate algorithm, though their algorithm does not probe the edges in random-order,
and attains its guarantee against a different fractional relaxation of the optimal probing algorithm. 

Our general guarantee can be parametrized by the patiences in the graph and improves in certain special cases.

\begin{corollary}\label{cor:432}
A $(1-\euler^{-2})/2\approx\textbf{0.432}$-approximate \textit{random-order} probing algorithm for general graphs without patience constraints.
\end{corollary}

In the special case where every patience is $\infty$, the guarantee of our random-order probing algorithm improves to 0.432.
We note that without patience constraints, the Greedy algorithm which probes edges in decreasing order of weights (ignoring the probabilities $p_e$) guarantees at least 1/2 the offline maximum weighted matching knowing the set of active edges in advance.
Even better guarantees are known if the edges can be probed in any order: \citet{gamlath2019beatingGreedy} derive a $(1-1/\euler)\approx0.632$-approximate algorithm for bipartite graphs by adding tightening constraints to the standard fractional relaxation; \citet{fu2021randomorder} derive a $\frac{8}{15}$-approximate algorithm for general graphs by considering random-order contention resolution with vertex arrivals, which was later improved slightly by \citet{macruryinduction2023}. These papers do not consider the finite patience case, and even in the infinite patience case, our algorithm differs because it probes the \textit{edges}
in a uniformly random order.
This means that the special case of our algorithm where patiences are $\infty$ implies results for \textit{Random-order Contention Resolution Schemes} and \textit{Prophet Secretary under Edge Arrivals}, as we discuss below.

\begin{corollary}\label{cor:matchingPolytope}
A $(1-\euler^{-2})/2$-balanced Random-order Contention Resolution Scheme for the matching polytope of general graphs.
\end{corollary}

For $c\le1$, the definition of a $c$-balanced Contention Resolution Scheme can be reduced to our goal of
probing every edge $e$ with probability at least $c\cdot y_e$, as we explain in \Cref{sec:2}.
Previous work in this area has derived a $(1-\euler^{-2})/2$-balanced ``offline'' Contention Resolution Scheme (CRS), which needs to know the set of active edges in advance, for the matching polytope of general graphs \citep{guruganesh2017understanding}.
The balancedness has since been shown to be strictly greater than $(1-\euler^{-2})/2$, while also satisfying a monotonicity property \citep{bruggmann2020optimal}.
However, both of these results require knowing the set of active edges in advance, while our algorithm observes the activeness of edges in an ordered fashion and must immediately decide whether to insert any active edges into the matching.
In fact, our algorithm satisfies the definition of a \textit{Random-order} Contention Resolution Scheme (RCRS) introduced in the papers by \citet{adamczyk2018random,lee2018optimal}. We note that 
the bound of $(1-\euler^{-2})/2$ has since been beaten by \citet{DBLP:journals/corr/abs-2205-08667}, and the state of the art is due to \cite{macrury_contention_2025}, the latter of which also shows that $1/2$ is unbeatable for an RCRS.

We note that the works of \citet{fu2021randomorder} and \citet{macruryinduction2023} also consider random-order contention resolution schemes for stochastic matching, but under \emph{vertex arrivals}, whereas our setting considers \emph{edge arrivals}. To this end, our specific setting of the matching polytope is not captured\footnote{
If the graph is bipartite, then its matching polytope can be captured by the intersection of two matroids; however, in this case the balancedness guaranteed by \citet{adamczyk2018random} is 1/3, which is worse than our balancedness of 0.432.
} by these papers. 

\begin{corollary}\label{cor:prophetSecretary}
A \textbf{0.432}-guarantee for the Prophet Secretary problem under Edge Arrivals, in general graphs.
\end{corollary}

Contention Resolution Schemes for the matching polytope also imply Prophet Inequalities under the Edge Arrival model introduced in \citet{gravin2019prophet}.
\citet{ezra2020online} attained a \textbf{0.337}-balanced \textit{adversarial-order} Contention Resolution Scheme for the matching polytope of general graphs which implies a $0.337$-guarantee for Prophet Inequalities under Edge Arrivals in any order. Recently, \citet{Nuti2026} achieved a $\frac{3 -\sqrt{5}}{2} \approx 0.3819$ guarantee on bipartite graphs, which is tight under certain concentration assumptions (see \cite{ma2024vanishing} for details). 
When the edges arrive in a uniformly random order, this can be called the \textit{Prophet Secretary} problem \citep{esfandiari2017prophet} under Edge Arrivals, for which our Random-order Contention Resolution Scheme implies an improved 0.432-guarantee.

\begin{corollary}\label{cor:rank1matroid}
A new class of non-adaptive $(1-1/\euler)$-balanced Random-order Contention Resolution Schemes for rank-1 matroids.
\end{corollary}

In the further special case of a star graph with infinite patiences, the balancedness of our probing algorithm improves to $1-1/\euler$.
The constraint that at most one edge in a star graph can be matched corresponds to a rank-1 matroid, for which a $(1-1/\euler)$-balanced Random-order Contention Resolution Scheme is already known \citep{lee2018optimal}.
However, our analysis yields a wide range of new such schemes, which are simpler than existing ones in that they satisfy a \textit{non-adaptiveness} property, which we we elaborate on in \Cref{sec:techniques}.

\begin{result}\label{result:unitPatience}
A $(1-1/\euler)\approx\textbf{0.632}$-approximate probing algorithm for bipartite graphs with a unit-patience side.
\end{result}

This result improves and generalizes the \textbf{1/3}-guarantee of \citet{hikima2021integrated} which holds for a special case of bipartite graphs with a unit-patience side. 
Despite the ubiquity of $(1-1/\euler)$-guarantees in online matching, to the best of our understanding, our guarantee requires the new technical ingredient of a $(1-1/\euler)$-balanced Ordered Contention Resolution Scheme for rank-1 matroids under
\textit{negative correlation}.
The aforementioned $(1-1/\euler)$-balancedness results for rank-1 matroids assume elements to be active independently and do not establish this guarantee, as we discuss in \Cref{sec:techniques}.
We note that our probing algorithm here must be able to choose the order, though.
We also note that \citet{borodin2025online} study an online problem which implies the same $(1-1/\euler)$-approximation for the offline problem in the particular case where one side has arbitrary patience and the other side has unlimited patience.

\subsection{Description of Techniques} \label{sec:techniques}\

\textbf{Random-order probing: finding an attenuation function which improves the worst case.}
As described before, our probing algorithm for general graphs considers the edges in a uniformly random order.
This can be implemented by each edge $e$ drawing an ``arrival time'' $x_e$ independently and uniformly from $[0,1]$.
\citet{baveja2018improved} have previously analyzed a similar algorithm, which probes each incoming edge $e$ according to the fractionally-feasible probability $y_e$ as long as $e$ is safe.
They show that every edge $e$ ends up being probed with probability at least $0.31\cdot y_e$, yielding a $0.31$-approximate algorithm.
The worst case occurs for an edge $e'$ whose values of $y_{e'},p_{e'}$ are close to 0, with both endpoints of $e'$ being incident to other edges $e''$ whose values of $y_{e''},p_{e''}$ are 1.

To improve this worst case, we attach an ``attenuation factor'' $a(e)\in[0,1]$ to each edge $e$ such that the probability of an incoming safe edge $e$ being probed is scaled down by a factor of $a(e)$.
We make $a(e)$ decreasing in $y_e$ and $p_e$, to dissuade the aforementioned edges $e''$ with large values of $y_{e''},p_{e''}$ from being probed and blocking edge $e'$.
However, given an arbitrary function $a$ defining the attenuation factors, computing the new worst case could be difficult.
Therefore, our approach is instead to derive properties on $a$ which cause the \emph{worst case to only involve edges $e$ with $a(e)=1$}.
More specifically, we show that for functions $a$ defined by $a(e)=\tilde{a}(y_ep_e)$
for some univariate function $\tilde{a}$ with $\tilde{a}(0)=1$, it is possible to design the derivatives of $\tilde{a}$ so that in the worst case, edge $e'$ is only incident to edges $e''$ with $y_{e''}p_{e''}\approx0$ (which implies that $a(e'')\approx \tilde{a}(0)=1$).
Our final bound is then derived via the same Poisson lower tail bound proven \citet{baveja2018improved} (namely \Cref{lem:poisson_lemma}), except that the worst case ratio has improved to 0.382. We note however that unlike the analysis of \citet{baveja2018improved}, the use of an attenuation function requires us to prove a non-trivial ``exchange argument'' in \Cref{lem:exchange_argument} of \Cref{sec:exchange_argument} when an endpoint of $e$ has patience $2$. The necessity of this computation was overlooked in the previous version of this paper, and so the analysis in this updated version is longer than before. We now make use of a related lemma and construction from \citet{derakhshan2026approximation}.

We note that the specific attenuation function we compute is inconsequential to this improved ratio---the key is showing the \emph{existence} of an attenuation function which \textit{eliminates} the previous worst case. However, by deriving such necessary properties, we are able to see not only which attenuation functions work, but also which \emph{don't work} (suggesting alternative attenuation functions can't do better than ours). Further, having a \emph{family} of valid attenuation functions allows for a choice of different attenuation functions for different application domains while keeping the same approximation guarantee.

\textbf{Using our attenuation for Random-order Contention Resolution Schemes.}
As stated in \Cref{cor:matchingPolytope,cor:prophetSecretary}, our attenuation analysis in the special case of infinite patiences implies a previously-unknown $(1-\euler^{-2})/2$-balanced Random-order Contention Resolution Scheme for the matching polytope of general graphs.
We now discuss the further special case in \Cref{cor:rank1matroid} of star graphs, for which we can contrast our technique with that used in the known Random-order Contention Resolution Schemes for rank-1 matroids.
Here, \citet{lee2018optimal} show that $(1-1/\euler)$-balancedness can be achieved using what we would call the attenuation function
$a(e)=\exp(-x_ep_e)$, which depends on the arrival time $x_e$ of each edge $e$.
This function is designed in \citep{ehsani2018prophet} to yield a closed-form expression for the probability of availability at any particular time $x\in[0,1]$, which allows them to elegantly compute that the balancedness is $1-1/\euler$.

In this special case, our analysis also yields a $(1-1/\euler)$-balanced Random-order Contention Resolution Scheme.
However, instead of designing a specific function, our analysis implies a class of functions which sufficiently\footnote{
This intuition can be illustrated as follows.
If there is no attenuation, then the worst case involves two edges $e',e''$ with probabilities $p_{e'}=0,p_{e''}=1$, which when shown in a random order implies that $e''$ will block $e'$ w.p.~1/2, whereas $e'$ will not block $e''$.  The goal of ``attenuation'' is to scale down the probability of selecting $e''$, to increase the probability of selecting $e'$.
} attenuate large values of $p_e$ to prevent them from blocking smaller values of $p_e$.
To elaborate, we show that the function $a(e)=\frac{1-p_e y_e}{1-\euler^{-(1-p_e y_e)}}(1-1/\euler)$ ensures that in the worst case, all edges $e$ have $p_e\approx0$.
And in this worst case, $a(e)\approx1$ for all $e$, from which we can conclude that any edge has an $\approx1-1/e$ chance of being selected.
We note that our function does not depend on the arrival time $x_e$ and can be seen as \textit{non-adaptive} $(1-1/e)$-balanced Random-order Contention Resolution Schemes for a rank-1 matroid, which we believe could be applied elsewhere.


\textbf{Bipartite graphs with a unit-patience side.}
For offline stochastic matching on bipartite graphs, a standard technique \citep{bansal2012lp} is to randomly round the fractionally-feasible values $y_e$ to binary values $Y_e$ using the dependent rounding procedure of \citet{gandhi2006dependent}, which ensures the patience constraints on both sides to be satisfied w.p.~1.
Under our additional assumption that one of the sides $V_{1}$ has unit-patience, the vertices in $V_{1}$ must have rounded degree at most 1, resulting in the rounded graph being a disjoint collection of stars.
One could then separately handle the edges in each star using a Random-order Contention Resolution Scheme for rank-1 matroids, since its edges $e$ will be disjoint from other stars and active independently w.p.~$p_e$.

However, this does not lead to $(1-1/\euler)$-balancedness.
To elaborate, for any vertex $v\notin V_1$, let $\delta(v)$ denote the set of edges incident to $v$.  Fractional feasibility of the $y_e$ values ensures
\begin{align} \label{eqn:intro}
\sum_{e\in\delta(v)}p_ey_e &\le1,
\end{align}
but the rounded star graph formed by edges $\{e\in\delta(v):Y_e=1\}$ could have $\sum_{e\in\delta(v)}p_eY_e>1$ whenever\footnote{
As a concrete example, let $v$ have patience $t_v=2$, and be incident to three edges with $p_1=1,y_1=\eps$; $p_2=1-\eps,y_2=1$; and $p_3=0,y_3=1-\eps$, which are fractionally feasible in that $\sum_{e\in\delta(v)}p_ey_e\le1$ and $\sum_{e\in\delta(v)}y_e\le2$.  The rounding must be such that whenever $Y_1=1$, we also have $Y_2=1$, which results in $\sum_{e\in\delta(v)}p_eY_e=2-\eps$.
} some particular edge $e'$ has $Y_{e'}=1$, making it difficult to ensure that edge $e'$ gets selected with sufficient probability when $Y_{e'}$ is rounded up.

\textbf{Reinterpretation as contention resolution under negative correlation.}
To resolve this issue, we instead imagine each edge $e\in\delta(v)$ as being active with probability $z_e:=p_ey_e$, which satisfies $\sum_{e\in\delta(v)}z_e\le1$, due to~\eqref{eqn:intro}.
The active edges in $\delta(v)$ are correlated in a way such that they cannot conflict with active edges in other stars; however, due to this correlation, any of the aforementioned Random-order Contention Resolution Schemes which are agnostic to the correlation will only be $1/2$-balanced, as we will show in \Cref{sec:unitPatience}.

Despite this apparent lack of a correlation-agnostic $(1-1/\euler)$-balanced Contention Resolution Scheme,
what we do show is that the optimal online algorithm, which trivially sorts the edges $e\in\delta(v)$ in decreasing order of weights $w_e$, obtains in expectation at least $1-1/\euler$ times the fractional value $\sum_{e\in\delta(v)}w_ez_e$.
This is only possible due to the following \textit{negative correlation} property enjoyed by the rounding procedure of \citet{gandhi2006dependent}, with $Z_e\in\{0,1\}$ denoting the activeness of an edge $e$:
\begin{align} \label{eqn:negCorr}
\Pr\left[\bigcap_{e\in S}(Z_e=b)\right] &\le\prod_{e\in S}\Pr[Z_e=b] &\forall S\subseteq\delta(v),b\in\{0,1\}.
\end{align}
Our analysis applies this negative correlation property with $b=0$ to show that the expected \textit{overall} weight obtained by the online algorithm is \textit{minimized when the $Z_e$'s are independent}, despite the fact that for a \textit{particular} edge $e'$, negative correlation among other edges in $\delta(v)$ \textit{could make it more likely that $e'$ is blocked} than in the independent case.
Through the equivalence derived in \citet{lee2018optimal}, our analysis also implies the existence of a $(1-1/\euler)$-balanced Ordered Contention Resolution Scheme for rank-1 matroids under negative correlation, assuming the order can be chosen. Some other recent papers have considered contention resolution schemes under various notions of negative dependence, such as \citet{dughmi2020outerlimits}, \citet{chekuri2021submodular}, and \citet{qiu2022submodular}. To our knowledge, it remains open
whether a $(1-1/\euler)$-balanced \textit{Random-order}\footnote{
A $1/2$-balanced Random-order Contention Resolution Scheme for star graphs under negative correlation is implied by the ``uniform black box'' in \citet{brubach2020attenuate}.  This black box has also been extended in some cases by \citet{fata2019multi}.
} Contention Resolution Scheme is possible under property~\eqref{eqn:negCorr}.

\subsection{Roadmap}
We begin in \Cref{sec:2} with some background. Then, in \Cref{sec:382Proof}, we present a random-order algorithm for general graphs, achieving our $0.382$-approximation for stochastic matching. We then analyze the same algorithm in the case of infinite patience, where the algorithm yields a $(1-\euler^{-2})/2$-approximation (and thus a $(1-\euler^{-2})/2$-balanced Random-order Contention Resolution Scheme for the matching polytope); and the case of a star graph, where the algorithm achieves an approximation guarantee of $1-1/\euler$ (and thus gives a $(1-1/\euler)$-balanced Random-order Contention Resolution scheme for rank-1 matroids). Finally, in \Cref{sec:unitPatience}, we present an algorithm for stochastic matching on bipartite graphs with a unit patience side, which achieves an approximation guarantee of $1-1/\euler$. 

\section{Notation and Preliminaries} \label{sec:2}

The weighted stochastic graph is denoted by $G=(V,E)$,
with the weight and probability of being active being denoted by $w_e$ and $p_e$, respectively, for each edge $e\in E$.
The patience parameter is denoted by $t_v$ for each vertex $v\in V$.
Given any problem instance defined by these values,
finding the optimal probing algorithm is computationally challenging \citep{bansal2012lp}.
For $c\le1$, a probing algorithm is said to be \textit{$c$-approximate} if its expected weight matched is at least $c$ times that of the optimal probing algorithm, for any problem instance.
The following LP relaxation is commonly used to derive computationally efficient probing algorithms.


\begin{subequations} \label{lp:matching}
  \begin{alignat}{3}
    \LP:=\max
    & \sum_{e\in E} w_{e}z_{e} \tag{\ref{lp:matching}} \\[2ex]
    \text{subject to }
    & \sum_{e\in \delta(v)} z_{e} \le 1 & &\hspace{3em}\forall v\in V \label{eq:lp-1match} \\
    & \sum_{e\in \delta(v)} y_{e} \le t_{v} & &\hspace{3em}\forall v\in V \label{eq:lp-tprobes} \\
    & 0 \le y_{e} \le 1 & &\hspace{3em}\forall e\in E \\
    & z_{e} = y_{e}p_{e} & & \hspace{3em}\forall e\in E
  \end{alignat}
\end{subequations}
Note that we have let $\delta(v)$ denote the set of edges incident to a vertex $v$.
The variable $y_{e}\in[0,1]$ corresponds to the probability of probing edge $e$.
The variable $z_{e}$ is then the probability that edge $e$ is included in the
matching (that is, it is both active and probed).
Constraint~\eqref{eq:lp-1match} for a vertex $v\in V$ is the \emph{matching
  constraint}: it is satisfied when $v$ is matched to at most one of its
neighbors in expectation. Constraint~\eqref{eq:lp-tprobes} for a vertex $v\in V$
is the \emph{patience constraint}: it is satisfied when at most $t_{v}$ edges
incident to $v$ are probed in expectation. We note that for a solution $(y_e)_{e \in E}$
to \eqref{lp:matching}, we denote $y(S) := \sum_{e \in S} y_e$ for $S \subseteq E$.

\begin{lemma}[\citet{bansal2012lp}] \label{lemma:ub}
For any problem instance, the optimal objective value $\LP$ is an upper bound on the expected weight matched by any optimal probing algorithm.
\end{lemma}

Due to \Cref{lemma:ub}, for an algorithm to be $c$-approximate, it suffices to show that its expected weight matched is at least $c\cdot\LP$.
All of our algorithms will be based on taking an optimal LP solution given by $(y_e)_{e\in E}$, and randomizing in a way to probe every edge $e$ with probability at least $c\cdot y_e$, which suffices for matching expected weight at least $c\cdot\LP$.
We note that the gap between the LP and the optimal probing algorithm can be large: \citet{brubach21FYS} show, via a result of \citet{karp1981maximum}, that for some graphs, the ratio between the maximum-weight matching and the LP objective value can be as large as $0.544$.

\begin{definition}[Ordered Contention Resolution Scheme Problem] \label{def:crs}
A graph $G=(V,E)$ and a vector $(\tilde{z}_e)_{e\in E}$ lying in its \textit{matching polytope} (i.e.\ satisfying $\sum_{e\in\delta(v)}\tilde{z}_e\le1$ for all $v$) is given.
Each edge $e\in E$ has an ``activeness'' in $\{0,1\}$ whose state is initially unknown other than that it equals 1 w.p.~$\tilde{z}_e$.
The activeness of edges is observed sequentially, and if an edge is both active and eligible to be matched (i.e.\ not incident to any edges already matched), then it can be either immediately matched or irrevocably discarded.
A (randomized) algorithm
that guarantees every edge $e\in E$ of being matched with ex ante probability at least $c\cdot \tilde{z}_e$ is said to be a 
\textit{$c$-balanced Ordered Contention Resolution Scheme} for the matching polytope.
\end{definition}

A probing algorithm which guarantees every edge $e$ probability at least $c\cdot y_e$ of being probed implies a $c$-balanced Ordered Contention Resolution Scheme.
To see this, given an instance to the problem in \Cref{def:crs}, we can construct an instance of offline stochastic matching with $p_e=\tilde{z}_e$ for all $e$ and $t_v=\infty$ for all $v$, which means that setting $y_e=1$ for all $e$ is a feasible solution to LP~\eqref{lp:matching}.
The probing algorithm will indicate whether to probe each edge in a way that guarantees the overall probability of any edge $e$ being probed to be at least $c\cdot y_e=c$.
Therefore, 
if in \Cref{def:crs} we ``accept an edge when active'' whenever the probing algorithm would have probed that edge, this translates to an ex ante guarantee of $c\cdot p_e=c\cdot \tilde{z}_e$ on the probability of any edge $e$ being matched, as desired.

\begin{definition}[Random-order]
A probing algorithm is said to be \textit{random-order} if it can be applied in the online setting where the edges arrive in a uniformly random order, and upon arrival, each edge needs to be either immediately probed (if safe) or irrevocably discarded.
Analogously, a \textit{Random-order Contention Resolution Scheme} must observe the activeness of edges in a uniformly random order.
\end{definition}

\section{Algorithm and Analysis for \Cref{result:382}}
\label{sec:382Proof}

\subsection{Description of Algorithm and Attenuation Functions}

Our algorithm is based on the algorithm of~\citet{baveja2018improved}, but with
an added attenuation factor. This algorithm first solves LP~\eqref{lp:matching}
to get a fractional solution $(y_{e})_{e\in E}$,
and independently draws $Y_{e} \in \{0,1\}$ for each $e \in E$,
where $\Pr[Y_e = 1] = y_e$. It generates a uniformly random \revisionchange{permutation on $E$}, and
processes the edges sequentially in the order of the permutation. 
Specifically, $e$ is probed if and only if $Y_{e}=1$ and both endpoints of $e$ are currently \textit{free} (i.e., when $e$ is processed in the permutation, its endpoints
are both \revisionchange{unmatched} and have remaining patience).

Our algorithm adds additional attenuation as follows. Let
$a\colon E\to [0,1]$
be our attenuation function. When we get to an edge $e=\{u,v\}$ in the permutation, we
draw a new independent Bernoulli random variable $A_{e}$
such that $\Pr[A_{e} =1] = a(e)$. Then, we probe $e$ if
\begin{enumerate}
\item $e$ is free,
\item $Y_{e} = 1$, and
\item $A_{e} = 1$
\end{enumerate}
Pseudocode is given in Algorithm~\ref{alg:attenuate}. Recall that we can
generate the uniformly random permutation by first independently drawing a
uniformly random ``arrival time'' $\pi(e)\in[0,1]$ for each $e \in E$, and ordering the edges in
increasing order of arrival times. 

\begin{algorithm}
  \caption{Attenuation-based algorithm for Stochastic Matching}
  \label{alg:attenuate}
  \begin{algorithmic}
    \Function{AttenuateMatch}{$V, E, \mathbf{p}$}
    \For{each edge $e$ in \emph{uniformly random order}}
    \State Generate random bit $Y_{e}=1$ with probability $y_{e}$
    \State Generate random bit $A_{e}=1$ with probability $a(e)$
    \If{$e \text{ is free}\land Y_{e}=1\land A_{e}=1$}
    \State Probe $e$
    \EndIf
    \EndFor
    \EndFunction
  \end{algorithmic}
\end{algorithm}
Our analysis allows for our $0.382$-approximation to be achieved for
many choices of attenuation function. Specifically, our analysis will require a
few key properties of our attenuation function, outlined in
\Cref{def:attenprops} below.
\begin{definition}
  \label{def:attenprops}
  We call an attenuation function $a(e)$ \emph{effective}
    if all of the following conditions hold:
  \begin{enumerate}
    \item $a(e)$ can be expressed as a function $\tilde{a}(z_{e})$ of $z_{e}$
    \item $\tilde{a}(0) = 1$
    \item $\ln(1-xz\tilde{a}(z))$ is a \emph{convex} function of $z\in[0,1]$ for any $x\in(0,1)$
  \end{enumerate}
\end{definition}

There are many functions which satisfy the conditions of \Cref{def:attenprops}. Notice, for instance, that the first two conditions are straightforward: we require only that the attenuation function be a function of $z_e$ and result in no attenuation when $z_e\approx 0$. The final condition is satisfied by many nice classes of functions, with some examples given in~\Cref{def:mainattenfunc}.
\begin{definition}
  \label{def:mainattenfunc}
  Define the following attenuation functions:
  \begin{itemize}
      \item The \emph{exponential} attenuation function, defined by $\tilde{a}_{\mathrm{exp}}(z) := \euler^{-\alpha z}$ for any $\alpha\ge 1/2$
      \item The \emph{linear} attenuation function, defined by $\tilde{a}_{\mathrm{lin}}(z) := 1 - \alpha z$ for any $\alpha\ge 1/2$
    \item The function $\tilde{a}(z) := \frac{1-z}{1-\euler^{-(1-z)}}(1-1/\euler)$
  \end{itemize}
\end{definition}
It can be easily verified, by taking second derivatives, that these functions
indeed satisfy the third property of \Cref{def:attenprops}, and hence are effective. Although
these examples are effective attenuation functions for all $\alpha\ge1/2$, in our
algorithms we will always set $\alpha=1/2$ to get the best bound at the end.

\subsection{Analysis of the Attenuation Algorithm} \label{sec:analysis_of_algorithm}
Suppose we execute \Cref{alg:attenuate} using an attenuation function $a: E \rightarrow [0,1]$.
For each $e \in E$, it will be convenient to denote
$P_e \in \{0,1\}$ as the indicator for whether $e$ is active, where $\Pr[P_e = 1] = p_e$.
Note that \Cref{alg:attenuate} decides to probe $e$ prior to learning $P_e$, and the event ``$e$ is free''
is independent of $Y_e$ and $A_e$. Thus, the expected weight of the matching
produced by \Cref{alg:attenuate} is 
\begin{align} \sum_{e\in E} w_{e} p_e  
\Pr[e\text{ is free}] \Pr[Y_{e}=1 \cap A_{e}=1]
= \sum_{e\in E}
w_{e}p_{e}y_{e} a(e) \Pr[e\text{ is free}]. \label{eq:reduction_to_edge}
\end{align}
%
Let us hereby assume that $a:E \rightarrow [0,1]$ is effective, and so $a(e) = \til{a}(p_e y_e)$ for each $e \in E$,
for some function $\tilde{a}: [0,1] \rightarrow [0,1]$ which satisfies \Cref{def:attenprops}.  Our goal is to lower bound $\tilde{a}(p_e y_e)\cdot \Pr[e\text{ is free} ]$ for all $e \in E$. Specifically,
define
\begin{equation}
    \beta^* :=
  \int_{0}^{1} \left(\euler^{-2x} + x\euler^{-2x}\right)^{2}dx \approx 0.38278.
\end{equation}
We prove the following theorem which holds for any solution $(y_e)_{e \in E}$ to \eqref{lp:matching}. Our main result then follows by taking $(y_e)_{e \in E}$ to be an optimal solution to \eqref{lp:matching}, and applying \eqref{eq:reduction_to_edge} and \Cref{lemma:ub}.
\begin{theorem}[corresponds to \Cref{result:382} from \Cref{sec:introResults}] \label{thm:rounding_guarantee}
    Suppose the stochastic graph $G=(V,E)$ has minimum patience $2$, and $(y_f)_{f \in E}$ is an arbitrary solution
    to the LP \eqref{lp:matching}. More, assume that $a: E \rightarrow [0,1]$ is taken from \Cref{def:attenprops}
    where $\alpha = 1/2$ if the exponential or linear attenuation function is chosen. Then, when \Cref{alg:attenuate}
    executes on $G$, 
    \begin{equation} \label{eqn:main_rounding}
      a(e) \cdot \Pr[\text{$e$ is free}] \ge  \beta^*,
        \end{equation}
    for all $e \in E$. 
\end{theorem}
\begin{remark} \label{rem:unit_patience}
If patience values of $1$ are allowed, there is a simple
input where the left-hand side of \eqref{eqn:main_rounding} is no greater than $1/3$ (no matter which attenuation function  $\tilde{a}$ with $\tilde{a}(0) =1$ is used). To see this, fix $\eps > 0$, and consider a single edge $e=(u,v)$ with $p_e =1$ and $y_e = \eps$, where $t_u = t_v = 1$. More, assume that $u$ and $v$ each are incident to one additional edge, say $f$ and $f'$, respectively,
    where $y_f = y_{f'} =1$, yet $p_{f}= p_{f'} = \eps$. Then, as $\eps \rightarrow 0$, there will then be no attenuation since $\tilde{a}(p_f y_f) = \tilde{a}(p_{f'} y_{f'}) = \tilde{a}(p_e y_e) = \tilde{a}(\eps) \rightarrow 0$. Moreover, the probability that $e$ is matched tends to $y_e/3$.
\end{remark}
\
\





In order to prove \Cref{thm:rounding_guarantee}, we provide a sufficient condition
for the event ``$e = \{u,v\}$ is free'' to occur. This sufficient condition
depends only on the random variables associated with the edges incident to the endpoints of $e$ -- i.e., $\delta(u) \cup \delta (v)$ -- which makes computing its probability tractable. Recall that each $f \in E$ has an arrival time $\pi(f) \in [0,1]$ drawn uniformly and independently. We say that $e = \{u,v\}$ is \textit{safe}
when executing \Cref{alg:attenuate} on $G$ with $(y_f)_{f \in E}$, provided that before $e$ is processed in the uniformly random permutation:
\begin{enumerate}
    \item At most $t_r -1$ probes were made to edges of $\delta(r)$ for each $r \in \{u,v\}$.
    \item For each $f \in \delta(u) \cup \delta(v) \setminus \{e\}$ with $\pi(f) < \pi(e)$, $ A_f P_f X_f = 0$.
\end{enumerate}
Observe that if $e$ is safe, then $e$ is free. Thus, using that $a(e) = \tilde{a}(p_e y_e)$,
\begin{equation} \label{eqn:original_output_probability}
    a(e) \cdot \Pr[\text{$e$ is free}]  \ge \tilde{a}(p_e y_e) \Pr[\text{$e$ is safe}]. 
\end{equation}
We next argue that with respect to lower bounding the right hand side of \eqref{eqn:original_output_probability},
we can assume our input is of a specific form:



\begin{lemma} \label{lem:noFloatingNew}
  Suppose \Cref{alg:attenuate} is executed using an effective attenuation function.
  Then, with respect to lower bounding the right hand side of \eqref{eqn:original_output_probability}, we may assume without loss of generality that:
  \begin{enumerate}
    \item $\sum_{f \in \delta(r)} y_f = t_r$ for each $r \in \{u,v\}$,
    \item $p_f \in \{0,1\}$ for all $f \in \delta(u) \cup \delta(v)$, and $p_e =1$.
  \end{enumerate}
\end{lemma}
\begin{myproof}

We focus on outlining why we may assume without loss that $p_f =1$ and $p_f \in \{0,1\}$ for all $f \in \delta(u) \cup \delta(v) \setminus \{e\}$ when lower bounding \eqref{eqn:original_output_probability}. It is clear we may assume $\sum_{f \in \delta(r)} x_f = t_r$ for each $r \in \{u,v\}$, as this can be attained
at the very end, where we add additional edges $f'$ incident to $r$ with $p_{f'}= 0$ and $y_{f'} = \min\{1, t_r - \sum_{f \in \delta(r)} y_f\}$, and then apply a simple coupling argument.

We first observe that since \eqref{eqn:original_output_probability} depends only on
the edges of $\delta(u) \cup \delta(v)$ for $e = \{u,v\}$, we may assume without loss that $E = \delta_G(u) \cup \delta_G(v)$ and $V = N_G(u) \cup N_G(v)$ (here $N_G(r)$ denotes the neighbors of $r$ in $G$ for $r \in \{u,v\}$). 

Under this assumption, we first show that we can assume $p_e =1$. To see this, construct $\bar{G}$, on the same edges and vertices of $G$, yet with LP solution $(\bar{y}_f)_{f \in E}$
and edge probabilities $(\bar{p}_f)_{f \in E}$, where $\bar{y}_e = p_e y_e$, $\bar{p}_e =1$
and $\bar{y}_f = y_f$, $\bar{p}_f = p_f$ for all $f \in E \setminus \{e\}$.
It is easy to check that $(\bar{y}_f)_{f \in E}$ is a feasible solution to
\eqref{lp:matching}. Moreover, $a$ is effective, and so $a(e) = \tilde{a}(\bar{p}_e \bar{y}_e) = \tilde{a}(p_e y_e)$. Finally, the probability that $e$ is safe is the same in either input, and so \eqref{eqn:original_output_probability} is the same in either input.

We now assume that $p_e = 1$ in the original input $G$.
It remains to argue that we can assume without loss that $p_f \in \{0,1\}$ for all $f \in E = \delta(u) \cup \delta(v) \setminus \{e\}$.
  We consider the ``worst-case structure'', i.e., the one that leads to the
  worst bound in \eqref{eqn:original_output_probability}. Suppose for contradiction that in this worst-case structure,
  there exists some edge $f\in E(u)$ with $p_{f}\in(0,1)$. We will ``split''
  this edge into two edges with integer probabilities and then see that this
  splitting can only \emph{decrease} the probability that $e$ is safe.

  We split the edge $f$ into two new edges $f_{0}$ and $f_{1}$ such that
  $p_{f_{0}} = 0$ and $p_{f_{1}}=1$. Then, set $y_{f_{0}} = y_{f}(1-p_{f})$ and
  $y_{f_{1}} = y_{f}p_{f}$. It is easy to see that this splitting does not affect any of the constraints
  of the LP and so is still feasible. This is because $E = \delta(u) \cup \delta(v)$, and $y_{f_{0}}+y_{f_{1}} = y_{f}$ and $y_{f_{0}}p_{f_{0}} + y_{f_{1}}p_{f_{1}} = y_{f}p_{f}$.

  Further, we argue that this splitting can only decreases the probability that
  $e$ is safe. Notice that if $e$ is not safe, then there must be some edge $f \in \delta(u) \cup \delta(v) \setminus \{e\}$ that is probed before $e$ and either: (a) is matched, or (b) causes $u$ to time out (i.e., it exhausts $u$'s patience by being the $t_{u}^{\text{th}}$ failed
probe); when this happens, we say that $f$ ``blocks'' $e$.

  We do this by showing that the probability of $e$ being blocked
  by the probing of $f$ is at most the probability of $e$ being blocked by the
  probing of either $f_{0}$ or $f_{1}$. Edge $f$ can block $e$ either by being
  successfully probed before $e$, or by an unsuccessful probe that causes $f$ to
  time out. 

  Let us condition on $\pi(e) = x$ for $x \in [0,1]$.
  First, we consider the (conditional) probability that $f$ blocks $e$ by a successful probe.
  This occurs with probability $xy_{f}a(f)p_{f}$. To see, note that $f$ occurs
  before $e$ with probability $x$; then, it is probed with probability
  $y_{f}a(f)$; finally, this probe is successful with probability $p_{f}$.

  Next, consider the probability that either $f_{0}$ or $f_{1}$ is successfully
  probed before $e$. This is
  $xy_{f_{1}}a(f_{1})p_{f_{1}} + xy_{f_{0}}a(f_{0})p_{f_{0}}$. Since
  $p_{f_{0}}=0$, this is simply
  $xy_{f_{1}}a(f_{1})p_{f_{1}} = xy_{f}p_{f}a(f_{1})$. Next, we use the fact
  that $a$ is effective, so $a(f) = \tilde{a}(z_{f})$. Since
  $z_{f} = y_{f}p_{f} = y_{f_{1}} = y_{f_{1}}p_{f_{1}} = z_{f_{1}}$, we have
  that $\tilde{a}(z_{f}) = \tilde{a}(z_{f_{1}})$ and so the probability is
  unchanged.

  Now consider the case of blocking $e$ by an unsuccessful probe (causing $e$ to
  time out). The probability that $f$ is unsuccessfully probed before $e$ is
  $xy_{f}a(f)(1-p_{f})$. Then, consider the probability that either $f_{0}$ or
  $f_{1}$ is probed unsuccessfully before $e$. For edge $f_{1}$, this occurs
  with probability $0$ since $p_{f_{1}}=1$. The probability that edge $f_{0}$ is
  unsuccessfully probed before $e$ is
  $xy_{f_{0}}a(f_{0})(1-p_{f_{0}}) = xy_{f_{0}}a(f_{0}) = xy_{f}a(f_{0})(1-p_{f})$.
  We again utilize the fact that $a$ is effective, noting that
  $a(f_{0}) = \tilde{a}(z_{f_{0}}) = \tilde{a}(0) = 1$. Thus, the probability of
  blocking $e$ by an unsuccessful probe is only increased by this splitting.

  All other edges are unchanged, so the overall probability of $e$ being safe
  only decrease by this splitting.
\end{myproof}

Let $E(r) := \delta(r) \setminus \{ e \}$ for $r \in \{u,v\}$. We also define
$E_{b}(r) = \{ f \in E(r) \mid p_f = b \}$ for $b \in \{0,1\}$. Given
\cref{lem:noFloatingNew}, we may assume that $p_e = 1$ and
$E(r) = E_0(r) \cup E_1(r)$ for each $r \in \{u,v\}$.

For $r \in \{u,v\}$, let us define $Q_0(r)$ as the number of queries
\Cref{alg:attenuate} makes to $E_0(r)$ before $e$ is processed. Observe then that 
if we condition on $\pi(e) = x$ for $x \in [0,1]$, then
\begin{align*}
  \Pr[\text{$e$ is safe} \mid \pi(e) = x]
  &= \Pr[\cap_{f \in E_{1}(u) \cup E_1(v)} \{ A_f Y_f \bm{1}_{\pi(f) < x} =0 \}  \mid \pi(e) = x] \prod_{r \in \{u,v\}} \Pr[Q_0(r) < t_r \mid \pi(e) = x]   \\
  &=  \prod_{f \in E_{1}(u) \cup E_{1}(v)} (1 - x a(f) y_f)  \prod_{r \in \{u,v\}} \Pr[Q_0(r) < t_r \mid \pi(e) = x].
\end{align*}
Using the the analytic properties of the attenuation function $\tilde{a}: [0,1] \rightarrow [0,1]$ from \Cref{def:attenprops}, we prove the following:
\begin{lemma} \label{lem:attenuation}
For each $r \in \{u,v\}$, we have that
    $\prod_{f \in E_{1}(r)} (1 - x a(f) y_f) \ge \euler^{-x  \sum_{f\in E_{1}(r)} y_{f}}$.
\end{lemma}
\begin{myproof}
Fix an arbitrary $r \in \{u,v\}$. Let us first define the follow quantities for convenience:
\[
  \text{$\lambda_1 := \prod_{f\in E_1(r)} (1-xy_fa(f))$ and $Y_1 := \sum_{f\in E_{1}(u)} y_{f}$}.
\]
  We consider $\ln \lambda_1 = \sum_{f\in E_{1}(r)} \ln(1 - xy_{f}a(f))$. Since
  $a$ is effective, it can be expressed as a function $\tilde{a}$ of $p_{f} y_f$.
  Notice that for $f\in E_{1}(r)$, $p_{f} =1$, and so we can write
  $a(f) = \tilde{a}(p_{f} y_f) = \tilde{a}(y_{f})$. Further, since $a$ is effective,
  we have that 
  $\ln(1 - xy_{f}\tilde{a}(y_{f}))$ is a convex function of $y_{f}\in[0,1]$ for any $x\in(0,1)$.

  Thus, the quantity $\ln\lambda_1$ is minimized by setting all $y_{f}$
  to be equal, i.e.\ $y_{f} = Y_1/|E_{1}(r)|$ for all $f\in
  E_{1}(r)$, and letting $|E_{1}(r)|$ tend to infinity. That is,
  \[
    \min\lambda_1 = \min \exp \left(\sum_{f\in E_1(r)}
      \ln(1-xy_f\tilde{a}(y_{f})) \right) = \lim_{k\to\infty}
    \left(1-\frac{x Y_1}{k}\tilde{a}\left(\frac{x Y_1}{k}\right)\right)^{k}
    = \euler^{-x Y_1},
  \]
  where the last line uses that $\tilde{a}(0) = 1$.
  Thus, for each $r \in \{u,v\}$,
  \[
    \prod_{f \in E_{1}(r)} (1 - x a(y_f) y_f) \ge \euler^{-x  \sum_{f\in E_{1}(r)} y_{f}}.
  \]
\end{myproof}
Recall that for any $S \subseteq E$, $y(S) := \sum_{f \in S} y_f$. Using
this notation, if we apply \Cref{lem:attenuation} to each $r \in \{u,v\}$, then
\[
 \prod_{f \in E_{1}(u) \cup E_{1}(v)} (1 - x a(y_f) y_f) \ge \euler^{-x(y(E_{1}(u)) + y(E_{1}(v)))},
\]
Thus, $\tilde{a}( y_e) \Pr[\text{$e$ is safe} \mid \pi(e) =x]$ is lower bounded
by
\begin{equation} \label{eqn:unsimplified_lower_bound}%
  \tilde{a}(y_e) \cdot \int_{0}^{1} \euler^{-x(y(E_{1}(u)) + y(E_{1}(v)))} \prod_{r \in \{u,v\}} \Pr[Q_0(r) < t_r \mid \pi(e) = x] dx.
\end{equation}
It remains to bound each term $\Pr[Q_0(r) < t_r \mid \pi(e) = x]$ for
$r \in \{u,v\}$. In order to do so, we require the following two lemmas. The
first is proven in \citet{baveja2018improved}, and the second is a standard
Chernoff bound.
\begin{lemma} [\citet{baveja2018improved}] \label{lem:poisson_lemma}%
  Fix an integer $t \ge 1$. Suppose $B_1, \ldots , B_k$ are independent
  Bernoulli's, and $\mu :=\sum_{i=1}^k \bE[B_i] < t - 1$. Then, if $\Pois(\mu)$
  is a Poisson random variable of parameter $\mu$,
  \[
    \Pr\left[\sum_{i=1}^k B_i < t \right] \ge \Pr[\Pois(\mu) < t] = \sum_{i=0}^{t -1} \frac{ \euler^{-x \mu} \mu^{i}}{i!}.
  \]

\end{lemma}
\begin{lemma} \label{lem:chernoff}%
  Fix $\eps > 0$. Suppose $B_1, \ldots , B_k$ are independent Bernoulli's, and
  $\mu :=\sum_{i=1}^k \bE[B_i]$. Then,
\[
\Pr\left[\sum_{i=1}^k B_i < (1+\epsilon)\mu \right] \ge 1 -\exp\left(-\frac{\epsilon^2}{2+\epsilon}\mu\right).
\]
\end{lemma}
To apply these bounds, note that
conditional on $\pi(e) = x$ for $x \in [0,1]$, $Q_{0}(r)$ is a sum of $|E_{0}(r)|$ Bernoulli random variables.
Moreover,
\[
\bE[ Q_{0}(r) \mid \pi(e) = x] = \sum_{f \in E_0(r)} a(p_f y_f) y_f x =  y( E_0(r)) x
\]
Thus, if $x \in [0, \frac{t_r -1}{y( E_0(r))})$,
then \Cref{lem:poisson_lemma} applies, and we may conclude that
\begin{equation} \label{eqn:poisson_application}
\Pr[Q_0(r) < t_r \mid \pi(e) = x] \ge  \Pr[ \Pois( y(E_0(r)) x) < t_r] = \sum_{i=0}^{t_r -1} \frac{ \euler^{-x y(E_0(r))} (x y(E_0(r)))^{i}}{i!}.
\end{equation}
On the other hand, for any $x \in [0,1]$, setting $\eps = (1-x)/x$
and $\mu = x y(E_0(r))$,
note that
\[
\frac{\mu \eps^2}{2 + \eps} = \frac{y(E_0(r)) (1 - x)^2}{1+x} \text{ and } (1+ \eps) x y(E_{0}(r)) = y(E_0(r)).
\]
Thus, we can apply \Cref{lem:chernoff} to conclude
that
\begin{equation*}
    \Pr[Q_0(r) < y(E_0(r)) \mid \pi(e) = x] \ge 1 - \exp\left( \frac{- y(E_0(r)) (1-x)^2}{1+x} \right)
\end{equation*}
However, since $t_r -1 \le y(E_0(r)) \le t_r$ by \Cref{lem:noFloatingNew}, we may combine the previous equation
with each bound on $y(E_0(r))$ to get that
\begin{equation} \label{eqn:chernoff_application}
    \Pr[Q_0(r) < t_r \mid \pi(e) = x] \ge 1 - \exp\left( \frac{-(t_r -1) (1-x)^2}{1+x} \right).
\end{equation}

We now apply \eqref{eqn:poisson_application} and \eqref{eqn:chernoff_application} to \eqref{eqn:unsimplified_lower_bound}
over $x \in [0,1]$, where which bound we use depends on the values of $t_u$ and $t_v$.
Due to the symmetry of the problem, we hereby assume that $t_u \le t_v$.
To eliminate a variable in the subsequent bounds,
we shall use that $y(E_0(r)) = t_r - y(E_1(r)) - y_e$ for each $r \in \{u,v\}$, as guaranteed by \Cref{lem:noFloatingNew}.
More, it will be convenient to define for $r \in \{u,v\}$,
\begin{equation} \label{eqn:x_r_critical}
x_{r,c} := (t_r -1)/(t_r - y(E_1(r)) - y_e),
\end{equation}
as well as $x_{c}:= \min\{x_{u,c},x_{v,c}\}$. Observe that \eqref{eqn:poisson_application} applies
to $\Pr[Q_0(r) < t_r \mid \pi(e) = x]$ for all $x \in [0, x_{r,c})$.

Next, we consider different cases for the patience: first, when $t_u \ge 3$, and then when $t_u = 2$.

\subsubsection{The case $t_u \ge 3$} \label{sec:mid_patience_u}



We first consider the case where $t_u \ge 3$. Let us further assume that
$\max\{t_u, t_v\} <48$. Then, by the definition of $x_c$, we can apply
\eqref{eqn:poisson_application} for $x \in [0, x_{c})$, followed by the trivial
lower bound of $0$ for $x \in [x_{c}, 1]$, and so
\eqref{eqn:unsimplified_lower_bound} is lower bounded by
\begin{align}
&\tilde{a}(y_e) \cdot \int_{0}^{x_c} \euler^{-x(y(E_{1}(u)) + y(E_{1}(v)))} \prod_{r \in \{u,v\}} \sum_{k=0}^{t_r -1} \frac{ \euler^{-x (t_r - y_e - y(E_1(r)))} (x (t_r - y_e - y(E_1(r)))^{k}}{k!} dy. \notag \\
&= \tilde{a}(y_e) \cdot \int_{0}^{x_c}\prod_{r \in \{u,v\}} \sum_{k=0}^{t_r -1} \frac{ \euler^{-x (t_r - y_e)} (x (t_r - y_e - y(E_1(r)))^{k}}{k!} dy. \label{eqn:middle_patience}
\end{align}
For each choice of $3 \le t_u \le t_v \le 47$, we numerically minimize \eqref{eqn:middle_patience} over variables $y_e$, $y(E_1(u))$ and $y(E_1(v))$,
subject to the constraints $y(E_1(r)) + y_e \le 1$
for each $r \in \{u,v\}$ (as guaranteed by \Cref{lem:noFloatingNew}).
The minimum of \eqref{eqn:middle_patience} occurs
when $t_u = t_v = 3$, $y_e = 0$, and $y(E_1(u)) = y(E_1(v)) = 1$, in which
case $x_{u,c} = x_{v,c} =1$, and so the value
$
  \int_{0}^{1} \left(\euler^{-3x} + 2x\euler^{-3x} + 2x^2 \euler^{-3x} \right)^2 dx \ge 0.385,
$
is attained, which is strictly greater than $\beta^*$.

Next, suppose that $t_u \le 47$, yet $t_v \ge 48$. In this case, we apply
\eqref{eqn:poisson_application} to $\Pr[Q_0(u) < t_u \mid \pi(e) = x]$ for
$x \in [0,x_{u,c})$, yet \eqref{eqn:chernoff_application} to
$\Pr[Q_0(v) < t_v \mid \pi(e) = x]$ for $x \in [0,1]$. After simplification,
this leaves us with the function
\begin{equation} \label{eqn:middle_high_patience}
\tilde{a}(y_e) \cdot \int_{0}^{x_{u,c}} \euler^{-xy(E_{1}(v))} \left(1 - \euler^{ \frac{-(t_v -1) (1-x)^2}{1+x}}\right) \sum_{k=0}^{t_u -1} \frac{ \euler^{-x (t_u - y_e)} (x (t_u - y(E_1(u)) - y_e))^{k}}{k!} dx,
\end{equation}
of variables $y_e, y(E_1(v))$, and $y(E_1(u))$ subject to the constraints
$y(E_1(r)) + y_e \le 1$ for $r \in \{u,v\}$. Now, for any $x \in [0,1]$,
$t_v \rightarrow \left(1 - \euler^{ \frac{-(t_v -1) (1-x)^2}{1+x}}\right)$ is an
increasing function in $t_v$, and $y(E_1(v)) \rightarrow \euler^{-xy(E_{1}(v))}$ is a
decreasing function in $y(E_1(v))$. Thus, the minimum of
\eqref{eqn:middle_high_patience} occurs when $t_v = 48$, and
$y(E_1(v)) = 1 - y_e$, in which case we are left with
\begin{equation} \label{eqn:middle_high_patience_simp}
  \tilde{a}(y_e) \cdot \int_{0}^{x_{u,c}} \euler^{-x(1 - y_e)}
  \left(1 - \euler^{ \frac{-47 (1-x)^2}{1+x}}\right) \sum_{k=0}^{t_u -1}
  \frac{ \euler^{-x (t_u - y_e)} (x (t_u - y(E_1(u)) - y_e))^{k}}{k!} dx.
\end{equation}
By numerically minimizing \eqref{eqn:middle_high_patience_simp} over $y_e$ and
$y(E_0(u))$ for each integer $t_u \in [3,47]$, the minimum occurs when
$t_u = 3$, $t_v = 48$, $y_e = 0$ and $y(E_1(u)) = 1$, in which case we get a
lower bound of
\[
  \int_{0}^1 \euler^{-x} \left(1 - \euler^{ \frac{-47 (1-x)^2}{1+x}}\right)
  \left(\euler^{-3x} + 2x\euler^{-3x} + 2x^2 \euler^{-3x} \right) dx
  \ge 0.384,
\]
which is again strictly larger than $\beta^*$.

Finally, when $t_u,t_v \ge 48$, we apply \eqref{eqn:chernoff_application} to
$\Pr[Q_0(r) < t_r \mid \pi(e) = x]$ for each $r \in \{u,v\}$ and all
$x \in [0,1]$. In this case, by again observing that the minimum occurs when
$t_u = t_v = 48$ and $y(E_1(u)) = y(E_1(v)) = 1 - y_e$, we are left with
\begin{equation}
    \tilde{a}(y_e) \cdot \int_{0}^{1} \euler^{-x(2 - 2y_e)} \left(1 - \euler^{ \frac{-47 (1-x)^2}{1+x}}\right)^2 dx,
\end{equation}
whose minimum is $\int_{0}^{1} \euler^{-2x} \left(1 - \euler^{ \frac{-47 (1-y)^2}{1+y}}\right)^2 dy \ge 0.392$ when $x_e = 0$, which is strictly greater than $\beta^*$.

\subsubsection{The case $t_u = 2$} \label{sec:exchange_argument}
When $t_u = t_v =2$, suppose that $y_e = 0$, and $y(E_1(u)) = 0$,
$y(E_1(v)) = 1$, in which case $y(E_0(u)) = 1$, $y(E_0(v)) = 2$ and so
$x_{u,c} =1/2$, $x_{v,c} =1 $ and $x_c = 1/2$. If we apply the bound of
\eqref{eqn:poisson_application} for $x \in [0,x_c]$ as done in
\eqref{eqn:middle_patience}, then the lower bound attained is
\begin{equation} \label{eqn:bad_range}
\int_{0}^{1/2} \left(\euler^{-2x} + \euler^{-2x} 2x \right)\left(\euler^{-2x} + \euler^{-2x} x \right)dx \approx 0.347,
\end{equation}
which is strictly less than $\beta^*$. In fact, even if we additionally apply
\eqref{eqn:chernoff_application} for $x \in [x_c, 1]$, the resulting bound is
still less than $\beta^*$. Similarly, using this analysis leads to bounds
strictly less than $\beta^*$ when $t_u =2$ and $t_v > 2$.

To get resolve this issue in the analysis, we revisit \eqref{eqn:unsimplified_lower_bound} for the case
when $t_u =2$, in which we have
\begin{equation} \label{eqn:unsimplified_lower_bound_restate}
  \tilde{a}(y_e) \int_0^1 \euler^{-x(y(E_{1}(u)) + y(E_{1}(v)))} \Pr[Q_0(u) < 2 \mid \pi(e) = x]
  \cdot  \Pr[Q_0(v) < t_v \mid \pi(e) = x] dx.
\end{equation}
Instead of directly applying \Cref{lem:poisson_lemma} to
\eqref{eqn:unsimplified_lower_bound_restate}, we shall first use an
\textit{exchange argument}, based on the construction from \citet{derakhshan2026approximation}
for the special case when $G$ is a star graph. Applied to our problem, this
involves constructing a supergraph of $\til{G}$ from $G$ by modifying
\textit{only} the neighborhood of $u$. In $\til{G}$, the analogously defined
term \eqref{eqn:unsimplified_lower_bound_restate} is no larger than that of $G$.
More, the new input $\til{G}$ will be constructed so that we can ultimately
apply \Cref{lem:poisson_lemma} over a larger range of $x \in [0,1]$, allowing us
to avoid the complication in \eqref{eqn:bad_range}.

We now provide the exact details. By applying the construction from
Lemma~3.10 of~\citet{derakhshan2026approximation} to $G$ and $(y_f)_{f \in E}$, we can modify the
neighborhood of $u$ to get a super-graph $\til{G} = (\til{V},\til{E})$ of $G$,
whose patience values, edges probabilities, fractional matching and arrival
times we denote by $(\til{t}_r)_{r \in \til{V}}$, $(\til{p}_f)_{f \in \til{E}}$,
$(\til{y}_f)_{f \in \til{E}}$ and $(\til{\pi}(f))_{f \in \til{E}}$, respectively.
This input has $t_r = \til{t}_r$ for each $r \in \{u,v\}$, satisfies the
properties of \Cref{lem:noFloatingNew}, leaves the neighboring edges of $v$ unchanged
(including $e$), and assigns the maximum fractional value to the edges
$f \in \delta_{\til{G}}(u)$ with $\til{p}_f =1$. Formally,
\begin{enumerate}
  \item $t_r = \til{t}_r$ for each $r \in \{u,v\}$, \label{enum:exchange_one}
  \item $\sum_{f \in \delta_{\til{G}}(r)} \til{x}_f = t_r$ for each
        $r \in \{u,v\}$, $\til{p}_{f} \in \{0,1\}$ for each
        $f \in \delta_{\til{G}}(e)$, and $\til{p}_e =1$.
  \item $\delta_{G}(v) = \delta_{\til{G}}(v)$, and $y_f = \til{y}_f$,
        $p_f = \til{p}_f$ for all $f \in \delta_{G}(v)$,
    \item $\sum_{f \in \delta_{\til{G}}(u): \til{p}_f =1} \til{y}_f = 1 - \til{y}_e$. \label{enum:exchange_last}
\end{enumerate}
Moreover, suppose that for each $r \in \{u,v\}$, we define $\til{E}_{q}(r):= \{f \in \delta_{\til{G}}(r) \setminus \{e\}: \til{p}_f = q\}$ for $q \in \{0,1\}$ and $\til{Q}_0(r)$ as the number of queries \Cref{alg:attenuate} makes to $\til{E}_0(r)$ before $e$ when executing on $\til{G}$ with $(\til{y}_f)_{f \in \til{E}}$. Then,
\begin{equation} \label{eqn:u_comp}%
  \int_{0}^1 \left(\euler^{-x y(E_{1}(u))} \Pr[Q_0(u) < 2 \mid \pi(e) = x] dy -\euler^{- x (1- y_e)} \Pr[\til{Q}_0(u) < 2 \mid \til{\pi}(e) = x] \right) dx \ge 0
\end{equation}
Let us denote $\til{y}(S) := \sum_{f \in S} \til{y}_f$ for each $S \subseteq \til{E}$.
We claim the following relation:
\begin{lemma} \label{lem:exchange_argument}
If $\til{G}$ and $(\til{y}_f)_{f \in \til{E}}$ are constructed from $G$ and $(y_f)_{f \in E}$ as above, then \eqref{eqn:unsimplified_lower_bound_restate} is lower bounded by
\begin{align}
  & \tilde{a}(\til{y}_e)\int_0^1 \euler^{-x(\til{y}(\til{E}_{1}(u)) + \til{y}(\til{E}_{1}(v)))}
    \Pr[\til{Q}_0(u) < \til{t}_u \mid \til{\pi}(e) = x] \cdot  \Pr[\til{Q}_0(v) < \til{t}_v \mid \til{\pi}(e) = x] dx. \label{eqn:unsimplified_lower_bound_til}
\end{align}
\end{lemma}


In order to prove \Cref{lem:exchange_argument}, it will be convenient to use the following elementary bound involving
the integral of a product of functions from \cite{macrury_contention_2025}.
\begin{proposition}[Proposition 11 in \cite{macrury_contention_2025}] \label{prop:integral}
Suppose that $\lambda, \phi:[0,1] \rightarrow \mathbb{R}$ are integrable, $\lambda \ge 0$, and $\lambda$ is non-increasing. Moreover, assume that there exists $0 \le z^* \le 1$ such that $\phi(z) \ge 0$ for all $z \in [0, z^*]$, and $\phi(z) \le 0$ for all $z \in [z^*,1]$. Then,
\[
    \int_{0}^{1} \lambda(z) \phi(z)  \, dz \ge \lambda(z^*) \int_{0}^{1} \phi(z) \, dz
\]
\end{proposition}

\begin{myproof}[Proof of \Cref{lem:exchange_argument}]
  We first simplify \eqref{eqn:unsimplified_lower_bound_til} using the
  properties \ref{enum:exchange_one}. to \ref{enum:exchange_last}. of $\til{G}$
  and $(\til{y}_f)_{f \in \til{E}}$ in relation to $G$ and $(y_f)_{f \in E}$.
  Specifically, $\til{t}_r = t_r$ for $r \in \{u,v\}$, $\til{y}_e = y_e$,
  $\til{y}(E_{1}(v)) = y(E_1(v))$, and $\til{y}(E_1(u)) = 1 - y_e$. Moreover, by
  coupling the two executions, we may also conclude that
  $\Pr[Q_0(v) < t_v \mid \pi(e) = x] = \Pr[\til{Q}_0(v) < \til{t}_v \mid \til{\pi}(e) = x]$
  for all $x \in [0,1]$. Thus, \eqref{eqn:unsimplified_lower_bound_til} is equal
  to
\begin{equation} \label{eqn:unsimplified_lower_bound_alt}
  \tilde{a}(y_e) \int_{0}^{1} \euler^{-x(1 - y_e + y(E_{1}(v)))} \Pr[\til{Q}_0(u) < 2 \mid
  \til{\pi}(e) = x] \cdot  \Pr[Q_0(v) < t_v \mid \pi(e) = x] dx.
\end{equation}
Now, subtracting \eqref{eqn:unsimplified_lower_bound_alt} from
\eqref{eqn:unsimplified_lower_bound_restate}, and dividing by $\tilde{a}(y_e)$, we are
left with
\small
\begin{align}
  \int_{0}^{1} \left(\euler^{-x y(E_{1}(u))} \Pr[Q_0(u) < 2 \mid \pi(e) = x] -
  \euler^{-x(1 - y_e)} \Pr[\til{Q}_0(u) < 2 \mid \til{\pi}(e) = x] \right)
  \euler^{-y(E_{1}(v))} \Pr[Q_0(v) < t_v \mid \pi(e) = x] dx \label{eqn:integral_comp}
\end{align}
\normalsize
The function $\lambda(x) := \euler^{-y(E_{1}(v))} \Pr[Q_0(v) < t_v \mid \pi(e) = x]$
is non-negative and non-increasing. Moreover, the function
$\phi(x):= \euler^{-x y(E_{1}(u))} \Pr[Q_0(u) < 2 \mid \pi(e) = x]- \euler^{-x(1 - y_e)} \Pr[\til{Q}_0(u) < 2 \mid \til{\pi}(e) = x]$
is initially non-negative and changes sign at most once on $[0,1]$. By combining
\Cref{prop:integral} with \eqref{eqn:u_comp}, we conclude that
\eqref{eqn:integral_comp} is non-negative, and so the lemma follows.
\end{myproof}
Due to \Cref{lem:exchange_argument}, with respect to lower bounding
\eqref{eqn:unsimplified_lower_bound_restate}, we may assume without loss that
the original input $G$ and fractional solution $(y_f)_{f \in E}$ satisfies
$y(E_0(u)) = 1 - y_e$. Recalling the definition of $x_{u,c}$ in
\eqref{eqn:x_r_critical}, this implies that $x_{u,c} = 1$. The remainder of the
proof depends on the value of $t_v$ and follows a similar structure to the
computations done in \Cref{sec:mid_patience_u}.

If $t_v =2$, then by the symmetry of $u$ and $v$, we may apply the exchange
argument of \Cref{lem:exchange_argument} to $v$ so as to ensure without loss
that $y(E_0(v)) = 1 - y_e$, and so $x_{v,c} =1$. If we then apply
\eqref{eqn:poisson_application} over all $x \in [0,1]$, we get a lower bound of
\begin{equation}
    \tilde{a}(y_e) \int_{0}^1 \left( \euler^{-x(2 - y_e)} + x\euler^{-x(2 - y_e)}\right)^2 dx,
\end{equation}
which is minimized when $y_e =0$, in which case we get a lower bound
of exactly $\beta^*$.

If $2 < t_v \le 47$, then since $x_{u,c} = 1$, we have
$x_{v,c} = x_c = (t_v -1)/(t_v - y(E_1(v)) - y_e)$. Thus, we get a lower bound
of
\begin{equation} \label{eqn:low_mid_patience}%
  \tilde{a}(y_e) \cdot \int_{0}^{x_{v,c}}( \euler^{-x (2- y_e)} + x\euler^{-x(2-y_e)}) \sum_{k=0}^{t_v -1}
  \frac{ \euler^{-x (t_v - y_e)} (x (t_v - y_e - y(E_1(v)))^{k}}{k!} dx,
\end{equation}
which is a function of variables $y_e$ and $y(E_1(v))$, subject to the
constraint $y(E_1(v)) + y_e \le 1$. We numerically verify that
\eqref{eqn:low_mid_patience} is minimized when $t_v =3$, $y_e = 0$ and
$y(E_1(v)) =1$, where it takes a value of
$\int_{0}^{1}( \euler^{-2x} + x\euler^{-2x})( \euler^{-3x} + x\euler^{-3x} + 2x^2 \euler^{-3x}) dx \ge 0.383$.

Finally, for $t_v \ge 48$, we apply \Cref{lem:chernoff} to get a lower bound of
\begin{equation} \label{eqn:low_high_patience}
\tilde{a}(y_e) \cdot \int_{0}^{1}( \euler^{-x (2- y_e)} + x\euler^{-x(2-y_e)}) \euler^{-xy(E_{1}(v))} \left(1 - \euler^{ \frac{-(t_v -1) (1-x)^2}{1+x}}\right) dx,
\end{equation}
which is a function of $y_e$ and $y(E_1(v))$,  subject to the constraint $y(E_1(v)) + y_e \le 1$.
By applying the same observations used to simplify \eqref{eqn:middle_high_patience},
we know that the minimum of \eqref{eqn:low_high_patience} occurs when $t_v = 48$
and $y(E_1(v)) = 1- x_e$. This leads to
\begin{equation} \label{eqn:low_high_patience_simp}
\tilde{a}(y_e) \cdot \int_{0}^{1}( \euler^{-x (2- y_e)} + x\euler^{-x(2-y_e)}) \euler^{-x(1-y_e)} \left(1 - \euler^{ \frac{-47 (1-x)^2}{1+x}}\right) dx,
\end{equation}
and we numerically verify that \eqref{eqn:low_high_patience_simp} attains its
minimum at $y_e =0$, where it is $\ge 0.3828$, a value strictly larger than
$\beta^*$.

\subsection{Infinite Patience}
When all patience values are $\infty$, the 
an edge cannot be blocked by any endpoint exhausting its patience, and so the lower bound on $a(e) \Pr[\text{$e$ is safe}]$ becomes $\tilde{a}(y_e)\int_{0}^{1} \euler^{-2x(1-y_e)} dx$. It is easy to verify
  that for the linear and exponential attenuation functions with $\alpha=1/2$,
the minimum still occurs at $y_{e}=0$. This gives us~\Cref{cor:432approx-main}.
\begin{corollary}[corresponds to \Cref{cor:432,cor:matchingPolytope,cor:prophetSecretary} from \Cref{sec:introResults}]
  \label{cor:432approx-main}
  In the case where all patiences are $\infty$, 
  for any feasible solution $(y_e)_{e\in E}$ to LP~\eqref{lp:matching}: \Cref{alg:attenuate}, using
  the effective attenuation functions from \Cref{def:mainattenfunc} with $\alpha=1/2$,
  considers the edges in a random order and probes every edge $e = \{u,v\}$ with probability at least
\begin{equation*}
    \left(\int_{x=0}^{1} \euler^{-2x} dx \right) y_e = \frac{1}{2}\left(1 - \euler^{-2} \right) y_e
\end{equation*}
which is $\approx0.432y_e$.
In this case, the algorithm is a $0.432$-approximation and a $0.432$-balanced Random-order Contention Resolution Scheme for the matching polytope.
\end{corollary}
Similarly, when all patience values are $\infty$, and we have a star graph, then the lower bound on $a(e) \Pr[\text{$e$ is safe}]$ simplifies to
$\tilde{a}(y_e)\int_{0}^{1} \euler^{-x(1-y_e)} dx$. In this case, taking $\tilde{a}(z) = \frac{1-z}{1-\euler^{-(1-z)}}(1-1/\euler) =  (1-1/\euler)/\int_{0}^{1} \euler^{-x(1-z)} dx$ gives an attenuation function, for which $\tilde{a}(y_e)\int_{0}^{1} \euler^{-x(1-y_e)} dx = 1-1/\euler$ for all $y_e \in [0,1]$.
\begin{corollary}
[corresponds to \Cref{cor:rank1matroid} from \Cref{sec:intro}]
\label{prop:stargraph}
For any star graph with infinite patiences and any feasible solution
$(y_e)_{e\in E}$ to the LP~\eqref{lp:matching}, Algorithm~\ref{alg:attenuate}, using the attenuation function
$\tilde{a}(z) = \frac{1-z}{1-\euler^{-(1-z)}}(1-1/\euler)$, considers the edges
in a random order and probes every edge $e=\{u,v\}\in E$ with probability at least
\begin{equation*}
\left(\int_{x=0}^{1} \euler^{-x} dx\right)y_e=\left(1-\frac{1}{\euler}\right)y_e.
\end{equation*}
This yields a $(1-1/e)$-balanced Random-order Contention Resolution Scheme for rank-1 matroids which does not adapt to the time of arrival of each element.
\end{corollary}

\section{Algorithm and Analysis for \Cref{result:unitPatience}}
\label{sec:unitPatience}

Let $G = (V, E)$ be a bipartite graph with bipartition $V=V_{1}\cup V_{2}$. Assume
$\pat_{u} = 1$ for all $u\in V_{1}$.

\textbf{Description of algorithm.}
We first solve the standard LP~\eqref{lp:matching} to obtain an optimal solution $(y_e)_{e\in E}$ satisfying
$\sum_{e\in \delta(v)} y_{e}p_{e} \le 1$ and
$\sum_{e\in \delta({v})}y_{e} \le \pat_{v}$ for all $v\in V$. Then, we run the rounding procedure of~\citet{gandhi2006dependent} on $y_{e}$ to get an integral solution $Y_{e}\in\set{0,1}$. This
guarantees that for each vertex $u\in V_{1}$ that $\sum_{e\in \delta(u)} Y_{e} \le 1$.
Thus, at most one vertex $e \in \delta(u)$ will be rounded to $Y_{e}=1$ for
every $u\in V_{1}$. Thus, in the rounded graph $\hat{G} := (V, \hat{E})$, 
where $\hat{E}:=\{e\in E:Y_e=1\}$, 
each vertex
$v\in V_{2}$ is the center of a star graph. For each vertex $v\in V_{2}$, we probe the
edges $e\in \delta(v)$ in decreasing order of weight, for each $e$ with
$Y_{e}=1$.

\textbf{Analysis of algorithm.}
The expected value achieved by this strategy is
\begin{equation*}
  \bE[\ALG] := \bE\left[\sum_{v\in V_2} W(v) \right] = \sum_{v\in V_2}\bE[W(v)]
\end{equation*}
where $W(v)$ denotes the weight of the edge matched by the algorithm (if any) for a vertex
$v$. 
In our analysis, we consider each vertex $v\in V_2$ separately, since in the rounded graph, it is the center of a star graph that is disconnected from any other vertices of $V_2$. We first utilize the negative correlation property of our dependent rounding technique~\citep{gandhi2006dependent} to establish the following lemma.
\begin{lemma} \label{lem:unitPatienceindbound}
Consider a fixed vertex $v\in V_2$. 
Label the edges of $\delta(v)$ from $1$ to $k := |\delta(v)|$ such that $w_1 \geq w_2 \geq \dots \geq w_k$. Then:
\begin{equation*}
    \bE[W(v)] \ge \sum_{i=1}^{k} w_{i} z_{i} \prod_{j=1}^{i-1} (1 - z_j)
\end{equation*}
\end{lemma}
\begin{myproof}
Recall from \Cref{sec:unitPatience} that $Y_{i}=1$ if edge $i$ was included in the rounding produced by the algorithm. Let $P_{i}$ be a Bernoulli random variable indicating whether edge $i$ is active. Then, we can write the expected value of $W(v)$ as
\begin{equation*}
  \bE[W(v)] = \bE\left[  \sum_{i=1}^{k}w_{i}Y_{i}P_{i}\prod_{i=1}^{i-1}(1-Y_{j}P_{j})\right]
\end{equation*}

Next, let $\hat{F}(i) = 1 - \prod_{j=1}^{i}(1-Y_{j}P_{j})$. Observe that
\begin{equation*}
  \hat{F}(i) = 1 - (1-P_{i}Y_{i})\prod_{j=1}^{i-1}(1-P_{j}Y_{j})
  = 1 - \prod_{j=1}^{i-1}(1-P_{j}Y_{j}) + P_{i}Y_{i}\prod_{j=1}^{i-1}(1-P_{j}Y_{j})
\end{equation*}
and thus we get
\begin{equation*}
  \hat{F}(i) - \hat{F}(i-1) = P_{i}Y_{i}\prod_{j=1}^{i-1}(1-P_{j}Y_{j}).
\end{equation*}
This gives the following expression for $W(v)$, letting $w_{k+1} := 0$ for convenience:
\begin{equation*}
  \bE \left[ \sum_{i=1}^{k}w_{i}(\hat{F}(i) - \hat{F}(i-1)) \right]
  = \bE \left[ \sum_{i=1}^{k}\hat{F}(i)(w_{i}-w_{i+1}) \right]
\end{equation*}
By the negative
correlation property~\eqref{eqn:negCorr}
of the dependent rounding~\citep{gandhi2006dependent}, and the independence of each $P_{j}$ from both $P_{j'}$ (for
all $j'\ne j$) and from $Y_{j'}$ for all $j'$, we can see that
\[
  \bE\left[ \prod_{j=1}^{i}(1-Y_{j}P_{j}) \right] \le \prod_{j=1}^{i}(1-y_{j}p_{j})
\]
and thus $\bE[\hat{F}(i)] \ge 1 - \prod_{j=1}^{i}(1-y_{j}p_{j})$. As usual, we let $z_{i} := p_{i}y_{i}$. We denote
$F(i) := 1 - \prod_{j=1}^{i}(1-z_{j})$, and from the negative
correlation~\eqref{eqn:negCorr}, we have that $\bE[\hat{F}(i)] \ge F(i)$, giving our desired result:
\begin{displaymath}
  \bE[W(v)] = \bE \left[ \sum_{i=1}^{k}\hat{F}(i)(w_{i}-w_{i+1}) \right]
  \geq \sum_{i=1}^{k} F(i)(w_{i} - w_{i+1})
  = \sum_{i=1}^{k}w_{i}z_{i}\prod_{j=1}^{i-1}(1-z_{j})
\end{displaymath}
\end{myproof}
Next, we work with the right-hand side expression, giving \Cref{lem:1minus1overe}.
\begin{lemma} \label{lem:1minus1overe}
  For $i\in[k]$, let $z_{i}\in[0,1]$ be such that
  $\sum_{i=1}^{k} z_{i} \le 1$. Let
  $w_{1}, w_{2}, \dots, w_{k} \in \bR^{+}$ be such that
  $w_{1} \ge w_{2} \ge \dots \ge w_{k}$. Then,
  \begin{equation} \label{eq:1infpat}
    \sum_{i=1}^{k}w_{i}z_{i} \prod_{j=1}^{i-1}(1-z_{j}) \ge
    \left(1-\frac{1}{\euler} \right)\sum_{i=1}^{k}w_{i}z_{i}
  \end{equation}
\end{lemma}
\begin{myproof}
  As in the previous proof, let $w_{k+1}=0$. Then, let
  $F(i) = 1 - \prod_{j=1}^{i}(1-z_{j})$ for $i\in\set{0,1,\dots,k}$. Note that
  $F(0) = 0$.

  We begin similarly to the proof of \Cref{lem:unitPatienceindbound}: Observe that
  \begin{equation*}
    F(i) = 1 - (1-z_{i})\prod_{j=1}^{i-1}(1-z_{j}) =
    1 - \prod_{j=1}^{i-1}(1-z_{j}) + z_{i}\prod_{j=1}^{i-1}(1-z_{j}) =
    F(i-1) + z_{i}\prod_{j=1}^{i-1}(1-z_{j})
  \end{equation*}
  and thus we get
  \begin{equation*}
    F(i) - F(i-1) = z_{i}\prod_{j=1}^{i-1}(1-z_{j}).
  \end{equation*}

  For convenience, define $w_{k+1}=0$. Then, we proceed as follows:
  \begin{equation} \label{eq:1infpatlhs}
    \sum_{i=1}^{k} w_{i}z_{i}\prod_{j=1}^{i-1}(1-z_{j}) = \sum_{i=1}^{k} w_{i}(F(i) - F(i-1)) = \sum_{i=1}^{k}F(i)(w_{i} - w_{i+1})
  \end{equation}

  Next, let $F^{*}(i) = \sum_{j=1}^{i} z_{j}$. It is easy to see then that
  $F^{*}(i) - F^{*}(i-1) = z_{i}$. Then we can observe that
  \begin{equation}
    \label{eq:1infpatrhs}
    \sum_{i=1}^{k}w_{i}z_{i} = \sum_{i=1}^{k}w_{i}(F^{*}(i) - F^{*}(i-1))
    = \sum_{i=1}^{k}F^{*}(i)(w_{i} - w_{i+1})
  \end{equation}

  Using the inequality of arithmetic and geometric means, we see that
  \begin{equation*}
    \prod_{j=1}^{i}(1 - z_{j}) \le \left( \frac{\sum_{j=1}^{i}(1-z_{j})}{i}\right)^{i}
    = \left(1 - \frac{\sum_{j=1}^{i}z_{j}}{i} \right)^{i}
    = \left(1 - \frac{F^{*}(i)}{i} \right)^{i} \le
    \euler^{-F^{*}(i)}
  \end{equation*}
  which gives us a bound on $F(i)$, shown in~\eqref{eq:Fi-lb}.
  \begin{equation}
    \label{eq:Fi-lb}
    F(i) \ge 1 - \euler^{-F^{*}(i)} \ge \left( 1 - 1/\euler \right)F^{*}(i)
  \end{equation}
  In~\eqref{eq:Fi-lb} above, we use the fact that
  $0\le F^{*}(i) = \sum_{j=1}^{i}z_{j} \le \sum_{j=1}^{k}z_{j} \le 1$. Then,
  combining~\eqref{eq:1infpatlhs}, \eqref{eq:Fi-lb}, and~\eqref{eq:1infpatrhs}
  gives the desired bound of~\eqref{eq:1infpat}.
  \begin{displaymath}
    \sum_{i=1}^{k} w_{i}z_{i}\prod_{j=1}^{i}(1-z_{j}) = \sum_{i=1}^{k}F(i)(w_{i} - w_{i+1}) \ge
    \sum_{i=1}^{k}(1-1/\euler)F^{*}(i)(w_{i} - w_{i+1}) = (1-1/\euler)\sum_{i=1}^{k}w_{i}z_{i}
  \end{displaymath}
\end{myproof}

For $v\in V_2$, let $\OPT(v) := \sum_{e\in\delta(v)} w_e z_e$. Using \Cref{lem:unitPatienceindbound}, we are then able to derive the following, from which our main result immediately follows.
\begin{lemma}
\label{lem:unitPatience}
    For any vertex $v\in V_2$, we have
    \[
        \bE[W(v)] \ge \left( 1 - \frac{1}{\euler} \right) \OPT(v)
    \]
\end{lemma}
\begin{myproof}
We can use \Cref{lem:unitPatienceindbound,lem:1minus1overe} together to get:
\[
\bE[W(v)] \ge \sum_{i=1}^{k} w_{i} z_{i} \prod_{j=1}^{i-1} (1 - z_j)
\ge \left(1-\frac{1}{\euler} \right)\sum_{i=1}^{k}w_{i}z_{i} = \left(1-\frac{1}{\euler} \right)\OPT(v)
\]
where the first inequality follows from \Cref{lem:unitPatienceindbound}, the second from \Cref{lem:1minus1overe}, and the final equality from the definition of $\OPT(v)$.
\end{myproof}
\begin{theorem}[corresponds to \Cref{result:unitPatience} from \Cref{sec:introResults}]
\label{thm:unitPatience}
For any bipartite graph with a unit-patience side, let $(y_e)_{e\in E}$ denote an optimal solution to the LP~\eqref{lp:matching}.
Then our algorithm above matches expected weight at least $(1-1/\euler)\sum_{v\in V_2}\sum_{e\in\delta(v)}w_ep_ey_e$, yielding a $(1-1/\euler)$-approximation.
\end{theorem}
\begin{myproof}[Proof of \Cref{thm:unitPatience}]
Applying \Cref{lem:unitPatience}, the expected value of our algorithm is
\begin{equation*}
    \bE[\ALG] = \sum_{v\in V_{2}}\bE[W(v)] \ge \sum_{v\in V_{2}}(1-1/\euler)\OPT(v) = (1-1/\euler)\OPT
\end{equation*}
as desired.
\end{myproof}

We note that the problem of stochastic matching on bipartite graphs with a unit patience side is a special case of particular interest, as it captures ``Problem~A'' in \citet{hikima2021integrated}. As shown in Lemma~A of~\citet{hikima2021integrated}, an $\alpha$-approximation for Problem~A implies an $\alpha$-approximation for the \emph{Integrated Stochastic Problem for Control Variables and Bipartite Matching} (ISPCB). This is discussed further at the end of this section.

\textbf{Why we cannot use an existing Contention Resolution Scheme.}
\Cref{lem:unitPatienceindbound,lem:unitPatience} show that the \textit{total} expected weight collected from edges in $\delta(v)$ is at least $(1-1/\euler)\cdot\OPT(v)$.
We now explain why it is not possible to use a correlation-agnostic Contention Resolution Scheme to match \textit{every} edge $e\in\delta(v)$ with probability at least $(1-1/\euler)$.
We consider the following example, in which every edge $e$ has the same value of $z_e=p_ey_e$.
Therefore, a correlation-agnostic Contention Resolution Scheme would treat the edges symmetrically, doing no better than a strategy which considers the edges in a random order until an edge $e$ is matched (which requires both $Y_e$ to be rounded to 1 and for edge $e$ to exist).
However, due to the first-stage dependent rounding for $Y_e$, the probabilities of edges being matched end up being negatively correlated (as defined in~\eqref{eqn:negCorr} in the introduction).
This negative correlation among the neighbors of a particular edge ``0'' can increase the probability of edge 0 being blocked to 1/2 (something not possible under independence), as we now demonstrate.

\begin{example}
Consider a star graph with $T+1$ edges, whose central vertex has patience 2.
Take the fractionally-feasible solution $y_0=1$, $p_0=1/(T+1)$, and $y_1=\cdots=y_T=1/T$, $p_1=\cdots=p_T=T/(T+1)$.
Any rounding procedure which satisfies the patience w.p.~1 and preserves the marginal probabilities will set $Y_0=Y_i=1$, $Y_{i'}=0$ for all other $i'$, with $i$ drawn uniformly from $\{1,\ldots,T\}$.
This implies that w.p.~$1-1/(T+1)$, one of the edges $1,\ldots,T$ will match upon being uncountered.
Any correlation-agnostic procedure would treat all edges symmetrically, since they all have the same value of $z_i=p_iy_i=1/(T+1)$.
Consequently, edge 0 will have a $(1-1/(T+1))/2$ probability of being blocked, whenever it is considered later than the aforementioned edge which matches upon being encountered.
Therefore, the correlation-agnostic procedure cannot be better than $1/2$-balanced as $T\to\infty$.
\end{example}

\subsection{Improvement to Approximation Ratio of~\citet{hikima2021integrated}}
The work of~\citet{hikima2021integrated} introduces and studies a problem they call \emph{Integrated Stochastic Problem for Control variables and Bipartite Matching} (ISPCB). This problem is a two-stage bipartite matching problem where the algorithm is given a bipartite graph $G = (V_1 \cup V_2, E)$ and must set a control variable $x_u$ for each $u\in V_1$. Then, each vertex $u\in V_1$ \emph{leaves} the graph with probability $p_u(x_u)$ (where $p_u(x_u)$ is some known probability which depends on $x_u$), and then the algorithm computes a maximum-weight matching on the resulting graph.

\citet{hikima2021integrated} prove an approximation guarantee for ISPCB by first studying a problem they denote Problem~A. Problem~A can be seen as the problem of stochastic bipartite matching with a unit-patience side which we study here: For each edge $(u,v) \in E$, $p_{uv} = p_u(x_u)$, $t_u = 1$ for each $u\in V_1$, and $t_v = \infty$ for each $v\in V_2$. The proof of Theorem~1 in~\citet{hikima2021integrated} shows that an $\alpha$-approximation of Problem~A implies an $\alpha$-approximation for ISPCB. The $1/3$-approximation then follows from the $1/3$-approximation for stochastic bipartite matching in~\citet{bansal2012lp}. Our \Cref{result:unitPatience} captures Problem~A, which thus implies a $(1-1/\euler)$-approximation for ISPCB, improving on the previous $1/3$-approximation of~\citet{hikima2021integrated}

\section*{Acknowledgments.} A preliminary version of this paper \citep{brubach2021improved} appeared at the 35th
Conference on Neural Information Processing Systems (NeurIPS), 2021. Nathaniel Grammel and Aravind Srinivasan were supported in part by NSF
award CCF-1749864, and by research awards from Amazon and Google.


\bibliographystyle{informs2014} 
\bibliography{bibliography} 

@article{brubach2021improved,
  title={Improved guarantees for offline stochastic matching via new ordered contention resolution schemes},
  author={Brubach, Brian and Grammel, Nathaniel and Ma, Will and Srinivasan, Aravind},
  journal={Advances in Neural Information Processing Systems},
  volume={34},
  year={2021}
}

@article{borodin2025online,
  title={Online bipartite matching in the probe-commit model},
  author={Borodin, Allan and MacRury, Calum},
  journal={Mathematical Programming},
  pages={1--54},
  year={2025},
  publisher={Springer}
}

@article{baveja2018improved,
  title={Improved bounds in stochastic matching and optimization},
  author={Baveja, Alok and Chavan, Amit and Nikiforov, Andrei and Srinivasan, Aravind and Xu, Pan},
  journal={Algorithmica},
  volume={80},
  number={11},
  pages={3225--3252},
  year={2018},
  publisher={Springer}
}

@article{fata2019multi,
  title={Multi-stage and multi-customer assortment optimization with inventory constraints},
  author={Fata, Elaheh and Ma, Will and Simchi-Levi, David},
  journal={Available at SSRN 3443109},
  year={2019}
}

@inproceedings{ma2024vanishing,
author = {Ma, Will and MacRury, Calum and Nuti, Pranav},
title = {Online Matching and Contention Resolution for Edge Arrivals with Vanishing Probabilities},
year = {2024},
isbn = {9798400707049},
publisher = {Association for Computing Machinery},
address = {New York, NY, USA},
url = {https://doi.org/10.1145/3670865.3673588},
doi = {10.1145/3670865.3673588},
booktitle = {Proceedings of the 25th ACM Conference on Economics and Computation},
pages = {159},
numpages = {1},
location = {New Haven, CT, USA},
series = {EC '24}
}

@unknown{Nuti2026,
author = {Aminian, Mohammad Reza and Niazadeh, Rad and Nuti, Pranav},
year = {2026},
month = {03},
pages = {},
title = {Stationary Online Contention Resolution Schemes},
doi = {10.2139/ssrn.6384199}
}

@inproceedings{macruryinduction2023,
  author       = {Calum MacRury and
                  Will Ma},
  editor       = {Bojan Mohar and
                  Igor Shinkar and
                  Ryan O'Donnell},
  title        = {Random-Order Contention Resolution via Continuous Induction: Tightness
                  for Bipartite Matching under Vertex Arrivals},
  booktitle    = {Proceedings of the 56th Annual {ACM} Symposium on Theory of Computing,
                  {STOC} 2024, Vancouver, BC, Canada, June 24-28, 2024},
  pages        = {1629--1640},
  publisher    = {{ACM}},
  year         = {2024},
  url          = {https://doi.org/10.1145/3618260.3649788},
  doi          = {10.1145/3618260.3649788},
  bibsource    = {dblp computer science bibliography, https://dblp.org}
}

@article{DBLP:journals/corr/abs-2205-08667,
  author       = {Tristan Pollner and
                  Mohammad Roghani and
                  Amin Saberi and
                  David Wajc},
  title        = {Improved Online Contention Resolution for Matchings and Applications
                  to the Gig Economy},
  journal      = {Math. Oper. Res.},
  volume       = {49},
  number       = {3},
  pages        = {1582--1606},
  year         = {2024},
  url          = {https://doi.org/10.1287/moor.2023.1388},
  doi          = {10.1287/MOOR.2023.1388},
  bibsource    = {dblp computer science bibliography, https://dblp.org}
}

@inproceedings{lee2018optimal,
  title={Optimal Online Contention Resolution Schemes via Ex-Ante Prophet Inequalities},
  author={Lee, Euiwoong and Singla, Sahil},
  booktitle={26th Annual European Symposium on Algorithms (ESA 2018)},
  year={2018},
  organization={Schloss Dagstuhl-Leibniz-Zentrum fuer Informatik}
}

@inproceedings{adamczyk2018random,
  title={Random order contention resolution schemes},
  author={Adamczyk, Marek and W{\l}odarczyk, Micha{\l}},
  booktitle={2018 IEEE 59th Annual Symposium on Foundations of Computer Science (FOCS)},
  pages={790--801},
  year={2018},
  organization={IEEE}
}

@incollection{adamczyk2015improved,
  title={Improved approximation algorithms for stochastic matching},
  author={Adamczyk, Marek and Grandoni, Fabrizio and Mukherjee, Joydeep},
  booktitle={Algorithms-ESA 2015},
  pages={1--12},
  year={2015},
  publisher={Springer}
}

@article{bansal2012lp,
  title={When LP is the cure for your matching woes: Improved bounds for stochastic matchings},
  author={Bansal, Nikhil and Gupta, Anupam and Li, Jian and Mestre, Juli{\'a}n and Nagarajan, Viswanath and Rudra, Atri},
  journal={Algorithmica},
  volume={63},
  number={4},
  pages={733--762},
  year={2012},
  publisher={Springer}
}

@inproceedings{hikima2021integrated,
  title={Integrated Optimization of Bipartite Matching and Its Stochastic Behavior: New Formulation and Approximation Algorithm via Min-cost Flow Optimization},
  author={Hikima, Yuya and Akagi, Yasunori and Kim, Hideaki and Kohjima, Masahiro and Kurashima, Takeshi and Toda, Hiroyuki},
  booktitle = {The Thirty-Fifth {AAAI} Conference on Artificial Intelligence, {AAAI} 2021}, 
  year={2021}
}

@inproceedings{ezra2020online,
  title={Online stochastic max-weight matching: prophet inequality for vertex and edge arrival models},
  author={Ezra, Tomer and Feldman, Michal and Gravin, Nick and Tang, Zhihao Gavin},
  booktitle={Proceedings of the 21st ACM Conference on Economics and Computation},
  pages={769--787},
  year={2020}
}

@article{guruganesh2017understanding,
  title={Understanding the correlation gap for matchings},
  author={Guruganesh, Guru and Lee, Euiwoong},
  journal={arXiv preprint arXiv:1710.06339},
  year={2017}
}

@article{bruggmann2020optimal,
  title={An optimal monotone contention resolution scheme for bipartite matchings via a polyhedral viewpoint},
  author={Bruggmann, Simon and Zenklusen, Rico},
  journal={Mathematical Programming},
  pages={1--51},
  year={2020},
  publisher={Springer}
}

@inproceedings{gravin2019prophet,
  title={Prophet inequality for bipartite matching: merits of being simple and non adaptive},
  author={Gravin, Nikolai and Wang, Hongao},
  booktitle={Proceedings of the 2019 ACM Conference on Economics and Computation},
  pages={93--109},
  year={2019}
}

@article{esfandiari2017prophet,
  title={Prophet secretary},
  author={Esfandiari, Hossein and Hajiaghayi, MohammadTaghi and Liaghat, Vahid and Monemizadeh, Morteza},
  journal={SIAM Journal on Discrete Mathematics},
  volume={31},
  number={3},
  pages={1685--1701},
  year={2017},
  publisher={SIAM}
}

@article{gandhi2006dependent,
  title={Dependent rounding and its applications to approximation algorithms},
  author={Gandhi, Rajiv and Khuller, Samir and Parthasarathy, Srinivasan and Srinivasan, Aravind},
  journal={Journal of the ACM (JACM)},
  volume={53},
  number={3},
  pages={324--360},
  year={2006},
  publisher={ACM New York, NY, USA}
}

@inproceedings{ehsani2018prophet,
  title={Prophet secretary for combinatorial auctions and matroids},
  author={Ehsani, Soheil and Hajiaghayi, MohammadTaghi and Kesselheim, Thomas and Singla, Sahil},
  booktitle={Proceedings of the twenty-ninth annual acm-siam symposium on discrete algorithms},
  pages={700--714},
  year={2018},
  organization={SIAM}
}

@article{brubach2020attenuate,
  title={Attenuate locally, win globally: Attenuation-based frameworks for online stochastic matching with timeouts},
  author={Brubach, Brian and Sankararaman, Karthik A and Srinivasan, Aravind and Xu, Pan},
  journal={Algorithmica},
  volume={82},
  number={1},
  pages={64--87},
  year={2020},
  publisher={Springer}
}

@inproceedings{DBLP:conf/aaai/Nanda0SDS20,
  author    = {Vedant Nanda and
               Pan Xu and
               Karthik Abinav Sankararaman and
               John P. Dickerson and
               Aravind Srinivasan},
  title     = {Balancing the Tradeoff between Profit and Fairness in Rideshare Platforms
               during High-Demand Hours},
  booktitle = {The Thirty-Fourth {AAAI} Conference on Artificial Intelligence, {AAAI}
               2020, The Thirty-Second Innovative Applications of Artificial Intelligence
               Conference, {IAAI} 2020, The Tenth {AAAI} Symposium on Educational
               Advances in Artificial Intelligence, {EAAI} 2020, New York, NY, USA,
               February 7-12, 2020},
  pages     = {2210--2217},
  publisher = {{AAAI} Press},
  year      = {2020}
  }

@inproceedings{DBLP:conf/ijcai/AhmadiADFK20,
  author    = {Saba Ahmadi and
               Faez Ahmed and
               John P. Dickerson and
               Mark Fuge and
               Samir Khuller},
  editor    = {Christian Bessiere},
  title     = {An Algorithm for Multi-Attribute Diverse Matching},
  booktitle = {Proceedings of the Twenty-Ninth International Joint Conference on
               Artificial Intelligence, {IJCAI} 2020},
  pages     = {3--9},
  publisher = {ijcai.org},
  year      = {2020}
  }

@inproceedings{DBLP:conf/aaai/XuSCDSSTT19,
  author    = {Pan Xu and
               Yexuan Shi and
               Hao Cheng and
               John P. Dickerson and
               Karthik Abinav Sankararaman and
               Aravind Srinivasan and
               Yongxin Tong and
               Leonidas Tsepenekas},
  title     = {A Unified Approach to Online Matching with Conflict-Aware Constraints},
  booktitle = {The Thirty-Third {AAAI} Conference on Artificial Intelligence, {AAAI}
               2019, The Thirty-First Innovative Applications of Artificial Intelligence
               Conference, {IAAI} 2019, The Ninth {AAAI} Symposium on Educational
               Advances in Artificial Intelligence, {EAAI} 2019, Honolulu, Hawaii,
               USA, January 27 - February 1, 2019},
  pages     = {2221--2228},
  publisher = {{AAAI} Press},
  year      = {2019}
  }

@inproceedings{DBLP:conf/ijcai/AhmedDF17,
  author    = {Faez Ahmed and
               John P. Dickerson and
               Mark Fuge},
  editor    = {Carles Sierra},
  title     = {Diverse Weighted Bipartite b-Matching},
  booktitle = {Proceedings of the Twenty-Sixth International Joint Conference on
               Artificial Intelligence, {IJCAI} 2017, Melbourne, Australia, August
               19-25, 2017},
  pages     = {35--41},
  publisher = {ijcai.org},
  year      = {2017}
  }

@InProceedings{brubach21FYS,
  title = 	 { Follow Your Star:  New Frameworks for Online Stochastic Matching with Known and Unknown Patience },
  author =       {Brubach, Brian and Grammel, Nathaniel and Ma, Will and Srinivasan, Aravind},
  booktitle = 	 {Proceedings of The 24th International Conference on Artificial Intelligence and Statistics},
  pages = 	 {2872--2880},
  year = 	 {2021},
  editor = 	 {Banerjee, Arindam and Fukumizu, Kenji},
  volume = 	 {130},
  series = 	 {Proceedings of Machine Learning Research},
  month = 	 {13--15 Apr},
  publisher =    {PMLR},
  url = 	 {http://proceedings.mlr.press/v130/brubach21a.html}
}

@inproceedings{esfandiari2016neuripsmatching,
 author = {Esfandiari, Hossein and Korula, Nitish and Mirrokni, Vahab},
 booktitle = {Advances in Neural Information Processing Systems},
 editor = {D. Lee and M. Sugiyama and U. Luxburg and I. Guyon and R. Garnett},
 pages = {},
 publisher = {Curran Associates, Inc.},
 title = {Bi-Objective Online Matching and Submodular  Allocations},
 url = {https://proceedings.neurips.cc/paper/2016/file/0966289037ad9846c5e994be2a91bafa-Paper.pdf},
 volume = {29},
 year = {2016}
}

@inproceedings{antoniadis2020neuripsmatching,
 author = {Antoniadis, Antonios and Gouleakis, Themis and Kleer, Pieter and Kolev, Pavel},
 booktitle = {Advances in Neural Information Processing Systems},
 editor = {H. Larochelle and M. Ranzato and R. Hadsell and M. F. Balcan and H. Lin},
 pages = {7933--7944},
 publisher = {Curran Associates, Inc.},
 title = {Secretary and Online Matching Problems with Machine Learned Advice},
 url = {https://proceedings.neurips.cc/paper/2020/file/5a378f8490c8d6af8647a753812f6e31-Paper.pdf},
 volume = {33},
 year = {2020}
}

@InProceedings{chen2009approximatingMatches,
author="Chen, Ning
and Immorlica, Nicole
and Karlin, Anna R.
and Mahdian, Mohammad
and Rudra, Atri",
editor="Albers, Susanne
and Marchetti-Spaccamela, Alberto
and Matias, Yossi
and Nikoletseas, Sotiris
and Thomas, Wolfgang",
title="Approximating Matches Made in Heaven",
booktitle="Automata, Languages and Programming",
year="2009",
publisher="Springer Berlin Heidelberg",
address="Berlin, Heidelberg",
pages="266--278",
isbn="978-3-642-02927-1"
}

@inproceedings{gamlath2019beatingGreedy,
author = {Gamlath, Buddhima and Kale, Sagar and Svensso, Ola},
title = {Beating greedy for stochastic bipartite matching},
year = {2019},
publisher = {Society for Industrial and Applied Mathematics},
address = {USA},
booktitle = {Proceedings of the Thirtieth Annual ACM-SIAM Symposium on Discrete Algorithms},
pages = {2841–2854},
numpages = {14},
location = {San Diego, California},
series = {SODA '19}
}

@inproceedings{karp1981maximum,
  title={Maximum matching in sparse random graphs},
  author={Karp, Richard M and Sipser, Michael},
  booktitle={22nd Annual Symposium on Foundations of Computer Science (sfcs 1981)},
  pages={364--375},
  year={1981},
  organization={IEEE}
}

@InProceedings{fu2021randomorder,
  author =	{Fu, Hu and Tang, Zhihao Gavin and Wu, Hongxun and Wu, Jinzhao and Zhang, Qianfan},
  title =	{{Random Order Vertex Arrival Contention Resolution Schemes for Matching, with Applications}},
  booktitle =	{48th International Colloquium on Automata, Languages, and Programming (ICALP 2021)},
  pages =	{68:1--68:20},
  series =	{Leibniz International Proceedings in Informatics (LIPIcs)},
  ISBN =	{978-3-95977-195-5},
  ISSN =	{1868-8969},
  year =	{2021},
  volume =	{198},
  editor =	{Bansal, Nikhil and Merelli, Emanuela and Worrell, James},
  publisher =	{Schloss Dagstuhl -- Leibniz-Zentrum f{\"u}r Informatik},
  address =	{Dagstuhl, Germany},
  URL =		{https://drops-dev.dagstuhl.de/entities/document/10.4230/LIPIcs.ICALP.2021.68},
  URN =		{urn:nbn:de:0030-drops-141376},
  doi =		{10.4230/LIPIcs.ICALP.2021.68}
}

@InProceedings{qiu2022submodular,
  author =	{Qiu, Frederick and Singla, Sahil},
  title =	{{Submodular Dominance and Applications}},
  booktitle =	{Approximation, Randomization, and Combinatorial Optimization. Algorithms and Techniques (APPROX/RANDOM 2022)},
  pages =	{44:1--44:21},
  series =	{Leibniz International Proceedings in Informatics (LIPIcs)},
  ISBN =	{978-3-95977-249-5},
  ISSN =	{1868-8969},
  year =	{2022},
  volume =	{245},
  editor =	{Chakrabarti, Amit and Swamy, Chaitanya},
  publisher =	{Schloss Dagstuhl -- Leibniz-Zentrum f{\"u}r Informatik},
  address =	{Dagstuhl, Germany},
  URL =		{https://drops-dev.dagstuhl.de/entities/document/10.4230/LIPIcs.APPROX/RANDOM.2022.44},
  URN =		{urn:nbn:de:0030-drops-171666},
  doi =		{10.4230/LIPIcs.APPROX/RANDOM.2022.44}
}

@InProceedings{dughmi2020outerlimits,
  author =	{Shaddin Dughmi},
  title =	{{The Outer Limits of Contention Resolution on Matroids and Connections to the Secretary Problem}},
  booktitle =	{47th International Colloquium on Automata, Languages, and Programming (ICALP 2020)},
  pages =	{42:1--42:18},
  series =	{Leibniz International Proceedings in Informatics (LIPIcs)},
  ISBN =	{978-3-95977-138-2},
  ISSN =	{1868-8969},
  year =	{2020},
  volume =	{168},
  editor =	{Artur Czumaj and Anuj Dawar and Emanuela Merelli},
  publisher =	{Schloss Dagstuhl--Leibniz-Zentrum f{\"u}r Informatik},
  address =	{Dagstuhl, Germany},
  URL =		{https://drops.dagstuhl.de/opus/volltexte/2020/12449},
  URN =		{urn:nbn:de:0030-drops-124496},
  doi =		{10.4230/LIPIcs.ICALP.2020.42}
}

@article{chekuri2021submodular,
  author       = {Chandra Chekuri and
                  Vasilis Livanos},
  title        = {On Submodular Prophet Inequalities and Correlation Gap},
  journal      = {CoRR},
  volume       = {abs/2107.03662},
  year         = {2021},
  url          = {https://arxiv.org/abs/2107.03662},
  eprinttype    = {arXiv},
  eprint       = {2107.03662},
  bibsource    = {dblp computer science bibliography, https://dblp.org}
}

@article{macrury_contention_2025,
author = {MacRury, Calum and Ma, Will and Grammel, Nathaniel},
title = {On (Random-Order) Online Contention Resolution Schemes for the Matching Polytope of (Bipartite) Graphs},
year = {2025},
issue_date = {March-April 2025},
publisher = {INFORMS},
address = {Linthicum, MD, USA},
volume = {73},
number = {2},
issn = {0030-364X},
url = {https://doi.org/10.1287/opre.2023.0339},
doi = {10.1287/opre.2023.0339},
journal = {Oper. Res.},
month = mar,
pages = {689–703},
numpages = {15}
}

@article{derakhshan2026approximation,
  title={Approximation Algorithms for Action-Reward Query-Commit Matching},
  author={Derakhshan, Mahsa and Ghasemi, Andisheh and MacRury, Calum},
  journal={arXiv preprint arXiv:2603.13487},
  year={2026}
}


\end{document}